\documentclass{elsarticle}
\usepackage{amsmath}
\usepackage{multirow}
\usepackage[table]{xcolor}

\usepackage{tabulary}
\usepackage{amssymb}
\usepackage{bm}
\usepackage{graphicx}
\usepackage{wasysym}
\usepackage{url}
\usepackage{algorithm}
\usepackage{algorithmic}
\usepackage{xcolor}
\usepackage{color}
\usepackage{subfig}
\usepackage{enumitem}
\usepackage{mathtools}
\usepackage{comment}
\usepackage{url}
\usepackage{tikz}
\usetikzlibrary{shapes,arrows,matrix,backgrounds,shapes,positioning,petri,topaths}
\usetikzlibrary{chains,calc}
\usepackage{pgfplots}
\pgfplotsset{compat=1.9}
\usepackage{pgfplotstable} %
\usepackage{booktabs}
\usepackage{pgf-umlsd}

\usetikzlibrary{shapes.geometric, arrows.meta, positioning}

\usepackage{eqparbox}

\usepackage{listings}

\lstdefinelanguage{AnBx}{
    keywords = {Protocol:, Types:, Agent, Number, Function, Certified, SymmetricKey, PublicKey, Untyped, Definitions:, Knowledge:, where, agree, share, Actions:, Goals:, weakly, authenticates, on, secret, between, empty},
    alsoletter=:,
    breaklines=true,
    basicstyle=\footnotesize\ttfamily,%
    keywordstyle=\bfseries,
    commentstyle=\color{green}\itshape,
    morecomment=[l]{\#}
}

\lstdefinelanguage{OFMCAttackTrace}{
    keywords = {Reached, State:,request,state_rB,state_rA,state_rs,state_rIntr,ATTACK, TRACE:},
    alsoletter=:,
    breaklines=true,
    basicstyle=\footnotesize\ttfamily,%
    keywordstyle=\bfseries,
    commentstyle=\color{green}\itshape,
    morecomment=[l]{\#}
}

\lstdefinestyle{mystyle}{
	basicstyle=\ttfamily\footnotesize,
	breakatwhitespace=false,         
	breaklines=true,                 
	captionpos=b,                    
	keepspaces=true,                 
	numbers=left,                    
	numbersep=5pt,                  
	showspaces=false,                
	showstringspaces=false,
	showtabs=false,                  
	tabsize=2
}

\newif\ifshowfixes \showfixestrue

\definecolor{pgreen}{rgb}{0,0.5,0}

\usepackage{nomencl}
\makenomenclature

\newcommand{\mytitle}{Evaluating Large Language Models for Symbolic Security Protocol Analysis}

\newcommand{\myabstract}{%
	Security protocol verification relies on formal tools such as ProVerif and OFMC. This study evaluates whether Large Language Models (LLMs) can perform comparable analysis. We test GPT and DeepSeek in chat and reasoning modes over three runs on 130 obfuscated AnB/AnBx protocols covering 388 security goals, scored against ProVerif and OFMC. Each provider uses a single model in both modes, switching reasoning on and off, so both contrasts isolate reasoning itself. Chat models achieve 72.7\% recall at 27.3\% precision for GPT and 69.3\% recall at 27.2\% precision for DeepSeek. Reasoning models reverse this trade-off, reaching 66.5\% precision and 54.5\% recall for GPT and 45.4\% precision and 57.3\% recall for DeepSeek. Enabling reasoning lifts precision from 27.3\% to 64.8\% for GPT and from 27.2\% to 44.4\% for DeepSeek on the consolidated verdict. The goal set is imbalanced, with 89 vulnerable goals against 299 secure ones; a trivial always-secure predictor scores 77.1\% accuracy, which only GPT reasoning exceeds. All models perform worst on authentication goals: reasoning models detect well under half of injective and non-injective agreement attacks, whereas chat models over-flag them at low precision. Confidentiality is the exception, with F1 up to 95.7\% in reasoning mode. Verdicts are unstable across runs: identical on 89.7\% of goals for GPT reasoning, 74.0\% for DeepSeek reasoning, 70.1\% for GPT chat, and 61.6\% for DeepSeek chat. Self-reported confidence is uniformly high yet shows no meaningful correlation with correctness. All results rest on a single zero-shot prompt and two model providers, which limits generalisability. On this benchmark, LLMs do not match formal verification, but may serve, at best, as pre-screening filters.
}

\newcommand{\mykeywords}{Security Protocols, Verification, Large Language Models}

\usepackage{boxedminipage}

\usepackage{ifpdf}
\usepackage{ifthen}

\ifpdf
  \newcommand{\inputfig}[1]{
  }
\else
  \newcommand{\inputfig}[1]{%
\fi

\newcommand{\tsigma}{\mathcal{T}_{\Sigma}}
\newcommand{\tsigmav}{\tsigma(\Var)}
\newcommand{\cryptn}[2]{\{#2\}_{#1}}
\newcommand{\sign}[2]{\mathsf{sign}_{#1}(#2)}
\newcommand{\scrypt}[2]{\{\!| #2 |\!\}_{#1}}
\newcommand{\inv}[1]{\iffont{inv}(#1)}
\newcommand{\invNA}{\iffont{inv}}
\newcommand{\pair}[2]{\langle #1, #2 \rangle}
\newcommand{\fst}[1]{\pi_1(#1)}
\newcommand{\snd}[1]{\pi_2(#1)}

\newcommand{\idfont}[1]{\mathit{#1}}
\newcommand{\HN}{\idfont{HN}}
\newcommand{\NA}{\idfont{NA}}
\newcommand{\NB}{\idfont{NB}}
\newcommand{\na}{\idfont{na}}
\newcommand{\nb}{\idfont{nb}}
\newcommand{\GX}{\idfont{GX}}
\newcommand{\GY}{\idfont{GY}}

\newcommand{\pkNA}{\iffont{pk}}
\newcommand{\skNA}{\iffont{sk}}
\newcommand{\pk}[1]{\pkNA(#1)}
\newcommand{\sk}[1]{\skNA(#1)}
\newcommand{\pkprime}[1]{\iffont{pk'}(#1)}

\newcommand{\senc}{\iffont{enc}}
\newcommand{\sdec}{\iffont{dec}}

\newcommand{\ckNA}{\mathsf{pk}}
\newcommand{\akNA}{\mathsf{sk}}
\newcommand{\ak}[1]{\akNA(#1)}
\newcommand{\ck}[1]{\ckNA(#1)}

\newcommand{\dyS}{\mathcal{DY}}
\newcommand{\dy}[1]{\dyS(#1)}
\newcommand{\dym}[2]{\dyS_{#1}(#2)}
\newcommand{\dymS}[1]{\dyS_{#1}}
\newcommand{\public}{\Sigma_p}

\newcommand{\state}[2]{\stateNA_{#1}(#2)}
\newcommand{\stateNA}{\iffont{state}}
\newcommand{\iknows}[1]{\iknowsNA(#1)}
\newcommand{\iknowsNA}{\iffont{iknows}}
\newcommand{\ifnot}[1]{\iffont{not}(#1)}
\newcommand{\ifdot}{{}_{\;\bullet\;}}

\newcommand{\ifarrow}[1][]{
  \ifthenelse{\equal{#1}{}}{
    \Rightarrow
  }{
    =\!\!\![{#1}]\!\!\!\hspace{1pt}\Rightarrow
  }}

\newcommand{\iffont}[1]{\mathsf{#1}}

\newcommand{\roleA}{\iffont{roleA}} %
\newcommand{\roleB}{\iffont{roleB}} %
\newcommand{\roleR}{\iffont{roleR}} %
\newcommand{\roleC}{\iffont{roleC}} %

\newcommand{\secCh}{\mbox{$\,\bullet\!\!\rightarrow\!\!\bullet\,$}}
\newcommand{\athCh}{\mbox{$\,\bullet\!\!\rightarrow$\,}}
\newcommand{\cnfCh}{\mbox{${\rightarrow\!\!\bullet\,}$}}
\newcommand{\secRCh}{\,\bullet\!\!\!\twoheadrightarrow\!\!\bullet\,}
\newcommand{\athRCh}{\,\bullet\!\!\!\twoheadrightarrow}
\newcommand{\insecCh}{\rightarrow}

\newcommand{\fresh}[1]{\stackrel{\mbox{\scriptsize @}\,}{#1}}
\newcommand{\forward}{\looparrowright}
\newcommand{\fforward}{\fresh{\forward}}
\newcommand{\farrow}{\fresh{\rightarrow}}

\newcommand{\onionCh}{\,\bullet[\rightarrow]\bullet\,}

\newcommand{\idmxCh}{\onionCh}

\newcommand{\dotChdot}{\,\bullet\!\!\leftrightarrow\!\!\bullet\,}
\newcommand{\dotCh}{\,\bullet\!\!\leftrightarrow}
\newcommand{\Chdot}{\leftrightarrow\!\!\bullet\,}

\newcommand{\PsecChP}{\,\circ\!\!\!\rightarrow\!\!\circ\,}
\newcommand{\PsecCh}{\,\circ\!\!\!\rightarrow\!\!\bullet\,}
\newcommand{\secChP}{\,\bullet\!\!\!\rightarrow\!\!\circ\,}
\newcommand{\PathCh}{\,\circ\!\!\!\rightarrow}
\newcommand{\cnfChP}{\rightarrow\!\!\circ\,}

\newcommand{\PChP}{\,\circ\!\!\leftrightarrow\!\!\circ\,}
\newcommand{\dotChP}{\,\bullet\!\!\leftrightarrow\!\!\dot\,}
\newcommand{\PChdot}{\,\circ\!\!\leftrightarrow\!\!\bullet\,}
\newcommand{\PCh}{\,\circ\!\!\leftrightarrow}
\newcommand{\ChP}{\leftrightarrow\!\!\circ\,}

\newcommand{\FathChNA}{\iffont{athCh}}
\newcommand{\FcnfChNA}{\iffont{cnfCh}}
\newcommand{\FsecChNA}{\iffont{secCh}}
\newcommand{\FfathChNA}{\iffont{fathCh}}
\newcommand{\FfsecChNA}{\iffont{fsecCh}}
\newcommand{\FsecCh}[4]{\FsecChNA(#1;#2;#3;#4)}
\newcommand{\FathCh}[3]{\FathChNA(#1;#2;#3)}
\newcommand{\FcnfCh}[2]{\FcnfChNA(#1;#2)}
\newcommand{\FfathCh}[4]{\FfathChNA^{#1}(#2;#3;#4)}
\newcommand{\FfsecCh}[5]{\FfsecChNA^{#1}(#2;#3;#4;#5)}

\newcommand{\athIssue}[1]{\iffont{athIssue(#1)}}
\newcommand{\cnfIssue}[1]{\iffont{cnfIssue(#1)}}
\newcommand{\secIssue}[1]{\iffont{secIssue(#1)}}
\newcommand{\athRely}[1]{\iffont{athRely(#1)}}
\newcommand{\cnfRely}[1]{\iffont{cnfRely(#1)}}
\newcommand{\secRely}[1]{\iffont{secRely(#1)}}

\newcommand{\pkEnc}[1]{\mathit{pkEnc}(#1)}
\newcommand{\pkSig}[1]{\mathit{pkSig}(#1)}

\newcommand{\DownGrade}{\mathrm{DownGrade}}
\newcommand{\Combine}{\mathrm{Combine}}
\newcommand{\SymOne}{\mathrm{Sym_1}}
\newcommand{\SymTwo}{\mathrm{Sym_2}}
\newcommand{\SymTre}{\mathrm{Sym_3}}
\newcommand{\CreatePseudo}{\mathrm{CreatePseudo}}
\newcommand{\UsePseudo}{\mathrm{UsePseudo}}
\newcommand{\UseRealName}{\mathrm{UseRealName}}
\newcommand{\PseudoDownGrade}{\mathrm{PseudoDownGrade}}
\newcommand{\AuthPseudo}{\mathrm{AuthPseudo}}
\newcommand{\AthTTP}{\mathrm{AthTTP}}
\newcommand{\CnfTTP}{\mathrm{CnfTTP}}
\newcommand{\SecTTP}{\mathrm{SecTTP}}

\newcommand{\honest}[1]{\mathit{honest}(#1)}
\newcommand{\dishonest}[1]{\iffont{dishonest}(#1)}
\newcommand{\vddash}{\vdash\!\!\!\vdash}
\newcommand{\mmodels}{\models\hspace{-0.3cm}\models}

\newcommand{\mode}[3]{({#1} \,|\, {#2} \,|\, {#3})}

\newcommand{\IK}{\mathit{IK}}
\newcommand{\agent}[1]{\iffont{agent}(#1)}

\newcommand{\Fresh}{\mathit{Fresh}}
\newcommand{\Payload}{\mathit{Payload}}
\newcommand{\Public}{\mathit{Public}}
\newcommand{\Tag}{\mathit{Tag}}
\newcommand{\lift}[1]{\lceil #1\rceil}

\newcommand{\pos}[1]{\mathit{pos}(#1)}
\newcommand{\Var}{\mathcal{V}}

\newcommand{\tagfont}[1]{\mathsf{#1}}
\newcommand{\tSone}{\tagfont{S_1}}
\newcommand{\tStwo}{\tagfont{S_2}}

\newcommand{\optfix}[2]{}
\newcommand{\dyM}[2]{\mathcal{DY}_{#1}(#2)}
\newcommand{\nf}[1]{#1_{\downarrow C/F}}
\newcommand{\pattern}[2]{{\lceil\,#1\,\rceil}_{#2}}
\newcommand{\decryptPat}[2]{{\lceil\!\!\lceil\,#1\,\rceil\!\!\rceil}_{#2}}

\newcommand{\pubNA}{\mathit{pub}}
\newcommand{\responseNA}{\mathit{response}}
\newcommand{\checkRNA}{\mathit{check}}

\newcommand{\pub}[1]{\pubNA(#1)}
\newcommand{\response}[1]{\responseNA(#1)}
\newcommand{\checkR}[1]{\checkRNA(#1)}

\newcommand{\Keyword}[1]{\mathsf{#1}}

\newcommand{\Protocol}[5]{
  \Keyword{Protocol:}~\mathit{#1}\\
  \Keyword{Types:}\\
  #2
  \Keyword{Knowledge:}\\
  #3
  \Keyword{Actions:}\\
  \begin{array}{cllllcl}
  #4
  \end{array}\\
  \Keyword{Goals:}\\
  \begin{array}{ccccl}
  #5
  \end{array}
}
\newcommand{\MidProtocol}[5]{
  \begin{array}{ccccl}
  #4
  \end{array}\\
}

\newcommand{\ShortProtocol}[5]{
  \Keyword{Protocol:}~\mathit{#1}\\
  \begin{array}{ccccl}
  #4
\end{array}\\
}
\newcommand{\CompactProtocol}[5]{
  \begin{array}{ccccl}
  #4
  \hline
  #5
  \end{array}\\
}
\newcommand{\Type}[2]{\quad #1\;\mathit{#2};\\}
\newcommand{\Agent}{\Keyword{Agent}}
\newcommand{\Number}{\Keyword{Number}}
\newcommand{\Function}{\Keyword{Function}}
\newcommand{\TFunction}{\Keyword{Function}}
\newcommand{\Knowledge}[2]{\quad\mathit{#1}:~\mathit{#2};\\}
\newcommand{\Create}[2]{
  \multicolumn{5}{l}{\quad\#\mathit{#1}~\text{creates}~\mathit{#2}}\\}
\newcommand{\Let}[2]{
  \multicolumn{5}{l}{\quad\#\mathit{#1}~:=~\mathit{#2}}\\}
\newcommand{\Action}[5]{\quad\mathit{#1}#2\mathit{#3},\mathit{#4}:&\mathit{#5}\\}
\newcommand{\Repeat}[5]{\multicolumn{5}{l}{
    \quad#1=\mathit{#2}#3\mathit{#4}:\mathit{#5}}\\}
\newcommand{\NGoal }[4]{\quad\mathit{#1}&#2&\mathit{#3}&:&\mathit{#4}\\}
\newcommand{\AuthGoal}[3]
{\multicolumn{5}{l}{
    \quad\mathit{#1}~\Keyword{authenticates}~\mathit{#2}~%
    \Keyword{on}~\mathit{#3}}\\}
\newcommand{\SecGoal}[2]
{\multicolumn{5}{l}{\quad\mathit{#1}~\Keyword{secret~between}~\mathit{#2}}\\}
\newcommand{\REML}[1]{\multicolumn{5}{l}{\quad\#\text{ #1}}\\}

\newcommand{\ifrule}[3]{#1\\ \ifarrow[#2]\\ #3\\[2ex]}

\newcommand{\subterm}{\sqsubseteq}
\newcommand{\propersubterm}{\sqsubset}
\newcommand{\supterm}{\sqsupseteq}
\newcommand{\propersupterm}{\sqsupset}

\newcommand{\decryptions}[2]{\mathit{decryptions}_{#1}(#2)}
\newcommand{\patternR}[3]{\pattern{#2}{#3}^{#1}}

\newcommand{\instanceof}{\succeq}

\newcommand{\secret}[2]{\iffont{secret}(#1,#2)}

\newcommand{\idemix}{\textsf{Identity Mixer}}

\newcommand{\xor}{\oplus}
\newcommand{\algo}[1]{\ensuremath{\mathsf{#1}}}
\newcommand{\const}[1]{\algo{#1}}
\newcommand{\vari}[1]{\ensuremath{\mathit{#1}}}

\newcommand{\ana}[2]{\mathit{ana}_{#1}(#2)}
\newcommand{\derivations}[3]{\mathit{derivations}_{#1}(#2,#3)}
\newcommand{\dereq}[3]{\mathit{dereq}_{#1}(#2,#3)}
\newcommand{\checks}[3]{\mathit{checks}_{#1}(#2,#3)}
\newcommand{\given}[1]{\textsl{given}(#1)}
\newcommand{\ungive}[1]{{#1}_*}

\definecolor{grey}{rgb}{0.5,0.5,0.5}\newcommand{\grey}{\color{grey}}
\definecolor{red}{rgb}{1,0,0}\newcommand{\red}{\color{red}}
\definecolor{green}{rgb}{0,0.45,0}\newcommand{\green}{\color{green}}
\definecolor{blue}{rgb}{0,0,1}\newcommand{\blue}{\color{blue}}

\newcommand{\rpif}{\mathbb{P}}
\newcommand{\rif}{\mathbb{R}}
\newcommand{\rf}{\mathbb{F}}
\newcommand{\re}{\mathbb{E}}
\newcommand{\md}{\mathbb{M}}
\newcommand{\traces}{\mathbb{T}}
\newcommand{\sigS}{\iffont{pk}(S)}
\newcommand{\send}[2]{\iffont{snd}_{#1}(#2)}
\newcommand{\recv}[2]{\iffont{rcv}_{#1}(#2)}
\newcommand{\trace}{t}
\newcommand{\evs}{\mathit{evs}}
\newcommand{\evt}{\mathit{evt}}
\newcommand{\mkset}[1]{[#1]} %
\newcommand{\player}[1]{\mathit{player}(#1)}
\newcommand{\used}[1]{{\mathit{ used\;#1}}}

\newcommand{\sem}[1]{[\![ #1 ]\!]}

\newcommand{\mX}{\mathcal{X}}

\newcommand{\PVar}{\mathcal{P}}
\newcommand{\SigmaP}{\Sigma_\PVar}
\newcommand{\tsigmap}{\mathcal{T}_{\SigmaP}}
\newcommand{\arity}[1]{\mathit{arity}(#1)}

\newcommand{\Nat}{\mathbb{N}}
\newcommand{\dcrypt}[2]{\cryptn{#1}{#2}^{-1}}
\newcommand{\dscrypt}[2]{\scrypt{#1}{#2}^{-1}}
\newcommand{\ccs}[1]{\mathit{ccs}(#1)}

\newcommand{\interpretation}{\mathcal{I}}
\newcommand{\intruder}{\mathsf{i}}

\newcommand{\know}[1]{\mathit{know}(#1)}
\newcommand{\verify}[1]{\mathit{verify}(#1)}
\newcommand{\true}{\mathit{true}}

\newcommand{\anl}[1]{&&&&\%\mathit{#1}\\}
\newcommand{\REM}[1]{\;\;\Keyword{\#}\;\;\text{#1}}

\newcommand{\StoreOA}[2]{\Keyword{Store}_{#1}(\,#2\,)}
\newcommand{\CheckStoreOA}[2]{\Keyword{CheckStore}_{#1}(\,#2\,)}
\newcommand{\LoadOA}[2]{#2\leftarrow\Keyword{Load}_{#1}}

\newcommand{\Store}[2]{\multicolumn{5}{l}{
    \quad\StoreOA{#1}{#2}}\\}
\newcommand{\Load}[2]{\multicolumn{5}{l}{
    \quad\LoadOA{#1}{#2}}\\}
\newcommand{\CheckStore}[2]{\multicolumn{5}{l}{
    \quad\CheckStoreOA{#1}{#2}}\\}

\newcommand{\FormNym}{\mathsf{FormNym}}
\newcommand{\GrantCred}{\mathsf{GrantCred}}
\newcommand{\VerifyCred}{\mathsf{VerifyCred}}
\newcommand{\VerifyCredOnNym}{\mathsf{VerifyCredOnNym}}
\newcommand{\masecNA}{\iffont{masec}}
\newcommand{\masec}[1]{\masecNA(#1)}
\newcommand{\ptagNA}{\iffont{p}}
\newcommand{\ptag}[1]{\ptagNA(#1)}

\newcommand{\commitNA}{\iffont{commit}}
\newcommand{\commitIINA}{\iffont{commit_2}}
\newcommand{\commitIIINA}{\iffont{commit_3}}

\newcommand{\commit}[1]{\commitNA(#1)}
\newcommand{\commitII}[1]{\commitIINA(#1)}
\newcommand{\commitIII}[1]{\commitIIINA(#1)}

\newcommand{\zkpNA}{\iffont{zkp}}
\newcommand{\user}[1]{\iffont{user}(#1)}
\newcommand{\owner}[1]{\iffont{owner}(#1)}

\newcommand{\Always}{\iffont{Always}}
\newcommand{\Previously}{\iffont{Previously}}

\newcommand{\pending}[1]{\iffont{pending}(#1)}

\newcommand{\pack}[1]{\mathit{pack}(#1)}
\newcommand{\hide}[1]{\mathit{hide}(#1)}

\newcommand{\ToTrusted}[1]{\Action{#1}{\idmxCh}{T}}
\newcommand{\TrustedTo}[1]{\Action{T}{\idmxCh}{#1}}

\newcommand{\registered}[1]{\iffont{registered}(#1)}
\newcommand{\granted}[1]{\iffont{granted}(#1)}

\newcommand{\verified}[1]{\iffont{verified}(#1)}

\newcommand{\dhh}{\mathit{dhh}}
\newcommand{\dhf}{\mathit{dhf}}
\newcommand{\DHtag}[1]{\mathit{tag}(#1)}
\newcommand{\hk}{\mathit{hk}}
\newcommand{\fk}{\mathit{fk}}
\newcommand{\exps}{\mathit{exps}}

\newcommand{\from}[2]{\fromnoarg(#1;#2)}
\newcommand{\fromnoarg}{\mathit{from}}

\renewcommand{\vector}[1]{\overline{#1}}

\newcommand{\vars}[1]{\mathit{vars}(#1)}
\newcommand{\dom}[1]{\mathit{dom}(#1)}

\newcommand{\drightarrow}{\rightarrow\!\!\!\!\!\rightarrow}
\newcommand{\bigsem}[1]{\left[\!\!\left[#1\right]\!\!\right]}

\newcommand{\hornarrow}{\rightarrow}
\newcommand{\iknowsa}[1]{\mathsf{iknows}(#1}
\newcommand{\signa}[1]{\mathsf{sign}_{#1}}
\newcommand{\todo}[1]{\fbox{\textbf{TODO: }#1}}
\newcommand{\valid}{\mathit{valid}}
\newcommand{\revoked}{\textit{revoked}}
\newcommand{\db}{\mathit{db}}
\newcommand{\ring}{\mathit{ring}}
\newcommand{\PK}{\mathit{PK}}
\newcommand{\NPK}{\mathit{NPK}}
\newcommand{\occurs}{\mathit{occurs}}
\newcommand{\timplies}{\mathit{implies}}
\newcommand{\LFP}{\mathit{LFP}}

\newcommand{\Habs}{\hat{\mathfrak{A}}}
\newcommand{\Abs}{\mathfrak{A}}
\newcommand{\Vara}{\Var_{\mathfrak{A}}}
\newcommand{\tabs}{\mathcal{T}_\mathfrak{A}}
\newcommand{\thabs}{\hat{\mathcal{T}_\mathfrak{A}}}
\newcommand{\ITIF}{{\it IF}}
\newcommand{\CryptoIF}{{\it CCM}}
\newcommand{\ChanIF}{{\it ICM}}

\newcommand{\ann}{\textit{ann}}
\newcommand{\atag}{\mathsf{atag}}
\newcommand{\fatag}{\mathsf{fatag}}
\newcommand{\ctag}{\mathsf{ctag}}
\newcommand{\stag}{\mathsf{stag}}
\newcommand{\fstag}{\mathsf{fstag}}
\newcommand{\btag}{\mathsf{plain}}
\newcommand{\ftag}{\mathsf{fresh}}
\newcommand{\fotag}{\mathsf{forw}}
\newcommand{\blind}{\mathsf{blind}}
\newcommand{\verif}[1]{#1}
\newcommand{\seenNA}{\mathsf{seen}}
\newcommand{\seen}[1]{\seenNA(#1)}
\newcommand{\notseen}[1]{\mathsf{fresh}(#1)}

\newcommand{\ttbracket}[1]{\texttt{[}#1\texttt{]}}
\newcommand{\anb}{\textit{AnB}}
\newcommand{\anbx}{\textit{AnBx}}
\newcommand{\IM}{\texttt{IM}}
\newcommand{\CM}{\texttt{CM}}
\newcommand{\trans}[3]{\llbracket#1\rrbracket^{#2}_{#3}}
\newcommand{\enc}[1]{\llbracket #1 \rrbracket}
\newcommand{\itrans}[3]{\llparenthesis #1 \rrparenthesis^{#2}_{#3}}
\newcommand{\after}[2]{#1 \texttt{ after } #2}
\newcommand{\arr}[1]{=\!\![#1]\!\!\Rightarrow}
\newcommand{\x}{{\cal X}}
\newcommand{\y}{{\cal Y}}
\newcommand{\X}{{\cal X}}
\newcommand{\T}{{\cal T}}
\newcommand{\V}{{\cal V}}
\newcommand{\chan}{\textsf{chan}}

\newcommand{\cchannel}{\mathsf{Chan_{CCM}}}
\newcommand{\ichannel}{\mathsf{Chan_{ICM}}}
\newcommand{\mgu}{\textit{mgu}}
\newcommand{\bl}{\mathsf{b}}

\renewcommand{\exp}{\iffont{exp}}

\renewcommand{\dscrypt}[2]{\scrypt{#1}{#2}}
\renewcommand{\dcrypt}[2]{\cryptn{#1}{#2}}

\newcommand{\signed}[2]{\mathsf{S}_{#1}(#2)}
\newcommand{\nonce}[1]{n_{#1}}
\newcommand{\newkey}[1]{\mathsf{new}\enskip{#1}}
\newcommand{\hmac}[2]{\mathsf{hmac}_{#1}(#2)}
\newcommand{\hashsymbol}{\mathcal{H}}
\newcommand{\hash}[1]{\mathsf{hash}(#1)}
\newcommand{\verdigest}[2]{[\textit{#1:#2}]}
\newcommand{\digest}[1]{[\textit{#1}]}

\newcommand{\contains}[2]{\iffont{contains}(#1,#2)}
\newcommand{\error}{\iffont{error}}
\newcommand{\wrequest}[1]{\iffont{wrequest}(#1)}
\newcommand{\request}[1]{\iffont{request}(#1)}
\newcommand{\witness}[1]{\iffont{witness}(#1)}

\newcommand{\maxd}[1]{\mathit{maxd}(#1)}
\newcommand{\sccs}[1]{\mathit{sccs}(#1)}
\newcommand{\nts}[1]{\mathit{#1}}

\newcommand{\channel}[2]{\mathsf{channel}_{#1}(#2)}
\newcommand{\GYS}{\mathit{GYS}}

\newcommand{\ccsF}[1]{\mathit{ccs}_F(#1)}

\newcommand{\olds}[1]{\oldstylenums{#1}}
\newcommand{\oldsb}[1]{{\bfseries\olds{#1}}}
\newcommand{\mnote}[1]{\stepcounter{ncomm}%
\vbox to0pt{\vss\llap{\tiny\oldsb{\arabic{ncomm}}}\vskip6pt}%
\marginpar{\tiny\bf\raggedright%
{\oldsb{\arabic{ncomm}}}.\hskip0.5em#1}}\newcounter{ncomm}

\newcommand{\miknote}[1]{{\red \mnote{{\red m: #1}}}}

\usepackage{hyperref}
\usepackage{doi}
\extrafloats{100}
\makeatletter
\def\ps@pprintTitle{%
	\let\@oddhead\@empty
	\let\@evenhead\@empty  \let\@oddfoot\@empty
	\let\@evenfoot\@oddfoot
}
\makeatother
\begin{document}
    \title{\mytitle}
	\author{Paolo Modesti}
	\ead{p.modesti@tees.ac.uk}
	\author {Syed Ahmed}
	\ead{D3776360@tees.ac.uk}
	\author{Ioannis Sfyrakis}
	\ead{i.sfyrakis@tees.ac.uk}
	\author {Derek Enodolomwanyi}
	\ead{E4481445@tees.ac.uk}
	\address{Teesside University, Middlesbrough, United Kingdom}

\begin{abstract}
\myabstract
\end{abstract}

 \maketitle

\section{Introduction}
\label{sec:introduction}

Security protocols govern how connected entities establish confidentiality, integrity, and authenticity during communication. Even when the underlying cryptographic primitives are sound, subtle flaws in protocol logic can expose entire systems to attack. The Needham--Schroeder public-key protocol, for instance, was considered secure for 17 years before Lowe demonstrated a man-in-the-middle attack that broke its authentication guarantees~\cite{10.1007/3-540-61042-1_43}.

Formal verification tools such as ProVerif~\cite{blanchet2016modeling} and OFMC~\cite{moedersheim2009open} provide mathematically rigorous analysis of protocol security properties under the Dolev--Yao attacker model~\cite{Dolev1983}. ProVerif translates protocol specifications into Horn clauses and can prove security for an unbounded number of sessions, whereas OFMC employs symbolic constraint solving for efficient detection of attack traces within a bounded session scope. These tools are highly effective, yet each carries practical limitations in termination, boundedness, or state-space explosion.

The recent emergence of Large Language Models (LLMs) capable of multi-step reasoning has sparked interest in whether such models can perform security analysis tasks that have traditionally been the preserve of formal methods. Chain-of-thought prompting and dedicated reasoning architectures have improved LLM performance on logical and mathematical tasks, raising the possibility of symbolic protocol analysis~\cite{Wang2026}. Nevertheless, serious concerns remain about the reliability of LLM outputs: hallucinations, overconfident incorrect verdicts, and high sensitivity to prompt phrasing have all been documented across application domains. Prior work~\cite{li2025constructing,mao2025llm,curaba2024cryptoformaleval} has focused on using LLMs to generate formal models or to translate protocol descriptions into structured representations, with correctness ultimately deferred to a dedicated verification tool. To the best of our knowledge, no study has yet systematically compared the security verdicts produced directly by an LLM against those of state-of-the-art formal verification tools while simultaneously contrasting chat-style and reasoning-oriented model configurations from different providers.

This paper addresses that gap through a comparative evaluation of OpenAI GPT and DeepSeek, each assessed in both chat and reasoning modes. Both pairs are like-for-like contrasts. Each provider is evaluated on a single model whose reasoning is switched off in chat mode and on in reasoning mode, so neither contrast is confounded by a change of model version (Section~\ref{sec:exp-setup}). We developed a Python pipeline that submits 130 obfuscated protocols, specified in the \anb{}/\anbx{} notation, to each model via their respective APIs, and collects structured JSON verdicts. Each verdict comprises a binary classification, either \emph{attack found} or \emph{no attack}, together with a confidence score, a textual justification, and an optional two-session attack trace. These outputs are evaluated against three formal sources: ProVerif, the primary benchmark owing to its unbounded session coverage, and OFMC at one and at two sessions. A waterfall priority scheme consolidates the three into a \emph{definitive} verdict covering all 388 goals, and we additionally report metrics against each source separately, using precision, recall, F1-score, and accuracy. The chat and the reasoning models were evaluated over three independent runs on the full dataset of 130 protocols and 388 security goals, with 95\% confidence intervals obtained from a cluster bootstrap that resamples protocols.

The work offers four main contributions:
\begin{enumerate}
	\item An automated pipeline for LLM-based security protocol analysis that integrates protocol obfuscation, a carefully engineered zero-shot prompt, and systematic evaluation against formal verification ground truth.
	\item A comparative analysis of chat and reasoning configurations, showing that the two fail in opposite ways. Chat models recall most attacks but hold precision below 29\%, whereas reasoning models reach up to 66.5\% precision and detect only just over half of the attacks. Each provider holds its model fixed across the two modes, so the gain is attributable to reasoning alone. On the consolidated verdict it lifts precision from 27.3\% to 64.8\% for GPT and from 27.2\% to 44.4\% for DeepSeek.
	\item An analysis of LLM confidence calibration, showing that self-reported confidence scores do not reliably distinguish correct from incorrect verdicts. Both models hold 94--99\% confidence on attacks they fail to detect.
	\item Empirical evidence that all evaluated models struggle with injective agreement, recovering 38.5\% of those attacks for GPT and 40.2\% for DeepSeek, and that they consistently miss attacks in specific protocol classes, thereby identifying practical settings in which LLM-based analysis remains unreliable.
\end{enumerate}

\paragraph{Outline of the paper} The paper is organised as follows. Section~\ref{sec:background} provides background on formal protocol verification, the \anb{}/\anbx{} notation, and the use of LLMs in security analysis. Section~\ref{sec:methodology} details the research design, model selection, prompt engineering, and evaluation metrics. Section~\ref{sec:implementation} describes the system architecture and the implementation of the analysis pipeline. Section~\ref{sec:evaluation} presents the experimental results, including a comparative analysis between chat and reasoning models, a breakdown by security goal type, an analysis of the persistent false negatives and false positives, an inter-run consistency assessment, a confidence calibration study, and a cost comparison. Section~\ref{sec:discussion} discusses the nature and reliability of LLM-generated security verdicts, the limitations of the work, and the practical implications. Section~\ref{sec:related} discusses the contribution in the context of the existing literature, and Section~\ref{sec:conclusion} summarises the findings and outlines future research directions.

\section{Background}
\label{sec:background}

\subsection{Formal Verification of Security Protocols}

The symbolic verification of security protocols relies on the Dolev--Yao adversary model~\cite{Dolev1983}, which abstracts cryptography as symbolic constructors and destructors under perfect cryptography. The intruder fully controls the network and may intercept, read, modify, inject messages, or initiate arbitrary concurrent sessions.

The intruder knowledge may be represented by a set of facts $\iknows{m}$, denoting that the intruder knows the term $m$, and the derivation capabilities are given by the closure rules shown in Figure~\ref{fig:Dolev-Yao-intruder-rules}. The function symbols of the algebra are partitioned into \emph{public} functions (available to the intruder) and \emph{private} functions (not available to the intruder, such as the inverse-key mapping $\inv{\cdot}$). Asymmetric encryption, including digital signing, is modelled through the operator $\cryptn{}{\cdot}$. Encryption is done with the public key of the intended recipient, and signature with the private key of the signer. Decryption or signature verification is possible only when the corresponding inverse key is known. Symmetric encryption is modelled by $\scrypt{}{\cdot}$, where the same key is used for both encryption and decryption. Tupling, projection, and application of public function symbols complete the deduction system. Within this model, a protocol is secure if the intruder can never derive a protected value (secrecy) or if the protocol guarantees certain correspondence properties across all admissible execution traces.

\begin{figure}[ht]
	\begin{centering}
		\begin{tabular}{rll>{\raggedright}m{3.3cm}}
			$\ensuremath{\iknows{M}.\iknows{K}}$ & $\ifarrow$ & $\iknows{\cryptn{K}{M}}$ & \multirow{3}{3.3cm}{\emph{Asymmetric}\\
				\emph{Encryption}}\tabularnewline
			\addlinespace
			$\iknows{\cryptn{K}{M}}.\iknows{\inv{K}}$ & $\ifarrow$ & $\iknows{M}$ & \tabularnewline
			\addlinespace
			$\iknows{\cryptn{\inv{K}}{M}}$ & $\ifarrow$ & $\iknows{M}$ & \tabularnewline
			\addlinespace
			$\ensuremath{\iknows{M}.\iknows{K}}$ & $\ifarrow$ & $\iknows{\scrypt{K}{M}}$ & \multirow{2}{3.3cm}{\emph{Symmetric}\\
				\emph{Encryption}}\tabularnewline
			\addlinespace
			$\iknows{\scrypt{K}{M}}.\iknows{K}$ & $\ifarrow$ & $\iknows{M}$ & \tabularnewline
			\addlinespace
			$\iknows{M}.\iknows{N}$ & $\ifarrow$ & $\iknows{M,N}$ & \emph{Tupling}\tabularnewline
			\addlinespace
			$\iknows{M,N}$ & $\ifarrow$ & $\iknows{M}.\iknows{N}$ & \emph{Projection}\tabularnewline
			\addlinespace
			$\iknows{M_{1}}.\cdots.\iknows{M_{n}}$ & $\ifarrow$ & $\iknows{f(M_{1},\ldots,M_{n})}$ & \emph{Function\\Application}\tabularnewline
			\addlinespace
		\end{tabular}
		\par\end{centering}
	\vspace{-4mm}
	\caption{\label{fig:Dolev-Yao-intruder-rules}Dolev-Yao intruder rules}
\end{figure}

Two mature tools that mechanise symbolic protocol analysis are ProVerif and OFMC, each with distinct strengths and limitations.

\paragraph{ProVerif}
ProVerif~\cite{blanchet2016modeling} translates protocol specifications, expressed in a variant of the applied pi-calculus, into a set of Horn clauses and applies a resolution-based algorithm to determine what the intruder can derive. It is notable for its ability to analyse protocols with an unbounded number of sessions, and it can verify secrecy, authentication (as correspondence properties), and some observational equivalences. ProVerif has been applied to real-world protocols such as TLS~\cite{blanchet2018composition} and e-voting systems~\cite{Cheval2023a}. Its abstraction, however, may introduce false attacks (positive approximations) and can sometimes lead to non-termination or state explosion on large, complex inputs.

\paragraph{OFMC}
The Open-source Fixedpoint Model Checker (OFMC)~\cite{10.1007/978-3-540-39650-5_15} operates on bounded session scenarios, typically one or two concurrent sessions. It combines lazy data-types for modelling infinite message structures with a lazy intruder technique that generates intruder knowledge only when required, enabling efficient symbolic constraint solving. OFMC is integrated into the AVISPA framework and accepts protocol descriptions in the \emph{Alice~\&~Bob} (\anb{}) \cite{DBLP:conf/IEEEares/Modersheim09} notation, making it a convenient tool for early design-time falsification. Its principal limitation is boundedness: a clean verification result applies only up to the chosen session bound, and attacks requiring more concurrent sessions may be missed. Moreover, OFMC may experience heavy memory consumption and state explosion with large protocols.

\vspace{2mm}
Both tools are sound with respect to their underlying models and produce deterministic results, but their practical deployment requires substantial expertise and computational resources. It is precisely these constraints that motivate the exploration of LLM-assisted protocol screening, where a fast, lightweight analysis, even if approximate, could help prioritise which protocols require full formal verification or expose fundamental design flaws.

\subsection{Protocol Specification in \anb{} and \anbx{}}

The \emph{Alice~\&~Bob} (\anb{}) \cite{DBLP:conf/IEEEares/Modersheim09} notation provides an intuitive, message-sequence oriented language for describing security protocols. Listing~\ref{fig:AnB-Protocol-Example} shows an \anb{} protocol that uses asymmetric encryption and a challenge–response mechanism to establish a fresh symmetric key and to exchange a confidential payload with mutual authentication. 

\begin{figure}[h!]
\begin{lstlisting}[
	language=AnBx,
	caption={\anb{} Protocol Example},
	label={fig:AnB-Protocol-Example}
	]
Protocol: Example_AnB
Types:
    Agent A,B;
    Number Msg,Nonce;
    Symmetric_key K;
    Function pk,sk,hash;
    Function log
Knowledge:
    A: A,B,pk,sk,inv(pk(A)),inv(sk(A)),hash,log;
    B: A,B,pk,sk,inv(pk(B)),hash,log
Actions:
    B -> A: {Nonce,B}pk(A)
    A -> B: {{Nonce,B,K}inv(sk(A))}pk(B)
    B -> A: {|Msg|}K
    A -> B: {|hash(Msg),log(A,Msg)|}K
Goals:
    K,Msg secret between A,B
    A authenticates B on Msg
    B authenticates A on K,Msg
    inv(pk(A)) secret between A
    inv(sk(A)) secret between A
\end{lstlisting}
\end{figure}

The abstract functions $\mathsf{pk}$ and $\mathsf{sk}$ model asymmetric encryption and signing keys, respectively, while $\mathsf{inv}$ is a private function mapping a public key to its associated private key, preventing the intruder from computing private keys from public ones. The goals of the example protocol include secrecy of the exchanged key and payload, as well as injective agreement on the payload between both parties, following the hierarchy of authentication goals introduced by Lowe~\cite{Lowe97}.

The \anbx{} language~\cite{jisa_2016} extends \anb{} with an explicit channel abstraction that captures the security guarantees expected of the underlying communication medium. \anbx{} supports the four channel types originally defined in \anb{}:
\begin{itemize}
	\item $\mathsf{A \insecCh B}$: an insecure channel, offering no protection.
	\item $\mathsf{A \cnfCh B}$: a confidential channel, guaranteeing that only the intended recipient can read the message.
	\item $\mathsf{A \athCh B}$: an authenticated channel, guaranteeing message origin without confidentiality.
	\item $\mathsf{A \secCh B}$: a secure channel, combining confidentiality and authentication.
\end{itemize}
\anbx{} also supports a richer notation for channels (specifying originators, verifiers, and intended recipients), including \emph{forwarding} channels that preserve security across intermediate agents, alongside private function declarations and user-definable macros. An \anbx{} specification (Listing~\ref{fig:AnBx-Protocol-Example}) can be compiled automatically into an equivalent \anb{} specification or into a formal model for ProVerif~\cite{anbx2015}, enabling cross-tool verification. An optional \emph{Definitions} section can be specified in \anbx{}, containing macros.

\begin{figure}[h!]
\begin{lstlisting}[
	language=AnBx,
	caption={\anbx{} Protocol Example},
	label={fig:AnBx-Protocol-Example}
	]
Protocol: Example_AnBx
Types:
    Agent A,B;
    Certified A,B;
    Number Msg;
    SymmetricKey K;
    Function [Agent,Number -> Number] log
Knowledge:
    A: A,B,log;
    B: B,A,log
Actions:
    A -> B, @(A|B|B): K
    B -> A: {|Msg|}K
    A -> B: {|hash(Msg),log(A,Msg)|}K
Goals:
    K,Msg secret between A,B
    A authenticates B on Msg
    B authenticates A on K,Msg
\end{lstlisting}
\end{figure}

Security goals in \anb{}/\anbx{} are classified into two broad families: authentication goals and secrecy goals.

\paragraph{Authentication goals}
Lowe's hierarchy~\cite{Lowe97} defines four authentication levels; we consider the two strongest. \emph{Non-injective agreement} guarantees that the two principals agree on a set of data values, but allows a single run of the responder to be paired with multiple initiator runs. \emph{Injective agreement} additionally requires a one-to-one correspondence between the initiator and responder runs, thereby ruling out replay attacks. Weaker forms (aliveness and weak agreement) are not examined here.

\paragraph{Secrecy goals}
Secrecy properties express that certain terms remain unknown to the Dolev--Yao intruder. \emph{Weak secrecy} holds if the intruder cannot derive the protected value from observed messages and allowed deduction steps. \emph{Strong secrecy} requires that the intruder cannot distinguish protocol executions in which the secret is replaced by an alternative value~\cite{Blanchet2023ProVerif}. This is the standard notion of secrecy in \anb{}/\anbx{}.

\subsection{Large Language Models for Security Analysis}

\emph{Large Language Models} (LLMs) are fundamentally next-token predictors trained on vast text corpora, yet through scaling and alignment they have acquired the ability to perform multi-step reasoning when guided by appropriate prompting strategies~\cite{achiam2023gpt}. \emph{Chain-of-thought} (CoT) prompting~\cite{Wei2022ChainOT} encourages the model to generate intermediate reasoning steps before producing a final answer, which has been shown to improve performance on arithmetic, commonsense, and symbolic tasks. More recently, dedicated \emph{reasoning models}, such as OpenAI’s o-series~\cite{openai2025o3} and DeepSeek-R1~\cite{guo2025deepseek}, have been trained with reinforcement learning to produce extended internal CoTs before emitting a response. These models have demonstrated state-of-the-art results on scientific and mathematical benchmarks~\cite{Wang2026}, raising the question of whether they can tackle symbolic security protocol verification.

Nevertheless, LLM outputs remain unreliable. Hallucinations, overconfident incorrect verdicts, and high sensitivity to prompt wording are well-documented across domains. In the security context, models may exhibit reasonable theoretical understanding but struggle to translate that understanding into correct, concrete analyses~\cite{curaba2024cryptoformaleval}. Furthermore, uncertainty in LLMs arises from multiple sources: input ambiguity, reasoning chain instability, parameter uncertainty, and prediction variability across runs~\cite{liu2025uncertainty}. This study operationalises the notion of uncertainty through a \emph{confidence} score (1–100) that the model is asked to self-report, allowing us to examine calibration quality.

Prior work on LLMs for cryptographic protocols has focused mostly on \emph{generating} formal models or translating informal descriptions into structured representations, leaving the soundness check to a dedicated tool. P2FGPT~\cite{li2025constructing} uses an LLM-based Generator–Checker–Modifier loop to produce ProVerif declarations from \anb{}-style specifications but reports that models can introduce errors while correcting previous ones. Mao et al.~\cite{mao2025llm} propose a staged pipeline for symbolic model generation, achieving correct outputs in 10 out of 18 cases. CryptoFormalEval~\cite{curaba2024cryptoformaleval} integrates an LLM agent with the Tamarin prover \cite{tamarin2012automated}, using Tamarin’s feedback to guide the agent’s analysis. In all these works, the LLM is an assistant to, not a substitute for, the formal verification tool. No study has yet treated the LLM as the primary verifier and directly compared its security verdicts against those of state-of-the-art formal tools across multiple model families and reasoning modes.

Effective prompt engineering is essential for structured tasks. In this work, we employ zero-shot prompting with a comprehensive system message that defines the cryptographic notation, security goal semantics, and a strict JSON output schema. This approach combines elements of in-context learning, through notation definitions, with constrained decoding using JSON format, balancing reasoning freedom with machine-parseable output. However, even carefully engineered prompts do not eliminate the fundamental uncertainty of LLM-generated security judgements, as we discuss in Section~\ref{sec:discussion}.

\section{Methodology}
\label{sec:methodology}

\subsection{Research Design}
We empirically compare LLM-generated security verdicts against formally verified ground truth. An experimental, multi-run design is employed because LLM behaviour is inherently probabilistic and sensitive to subtle prompt variations. A single run cannot adequately characterise performance~\cite{achiam2023gpt,Wei2022ChainOT}. By executing each model configuration several times over the same set of protocols, we obtain both per-run classification metrics and inter-run consistency statistics, enabling an assessment of the stability and repeatability of LLM-based protocol analysis.

\subsection{Model Selection}
We selected two LLM providers, OpenAI and DeepSeek, to capture differences in training data, model architecture, and optimisation strategy. Each provider was evaluated in two operational modes:
\begin{itemize}
	\item \emph{Chat mode}: \texttt{gpt-5.4} \cite{OpenAI2026APIPricing} with its reasoning effort set to \texttt{none}, and \texttt{deepseek-v4-flash} run in non-thinking mode, generate answers directly, without an explicit intermediate reasoning phase.
	\item \emph{Reasoning mode}: the same two models, \texttt{gpt-5.4} at reasoning effort \texttt{high} and \texttt{deepseek-v4-flash} with thinking enabled, produce extended internal chain-of-thought traces before returning a final verdict, a capability that has yielded substantial gains on logical and mathematical benchmarks \cite{openai2025o3,guo2025deepseek}.
\end{itemize}
Each provider therefore contributes a matched pair: one model, one prompt, and a single controlled difference, namely whether the model reasons before answering. A difference between the two modes can then be attributed to reasoning rather than to a change of model version, for both providers alike. Comparing these modes tests the hypothesis that explicit, multi-step reasoning improves accuracy on symbolic protocol verification tasks.

Two deliberate scoping choices bound what this design can show. The study uses a single zero-shot prompt, identical across all four configurations, and two model providers. Holding the prompt fixed is what makes the mode contrast interpretable, but it also means every result reported here is conditional on that one prompt. Prompt sensitivity is a well-documented property of LLMs \cite{white2023prompt,bae2024enhancing}. Section~\ref{sec:limitations} returns to both constraints.

\subsection{Dataset and Protocol Obfuscation}\label{subsec:obfuscation}
The evaluation dataset consists of 130 security protocols specified in \anb{}/\anbx{} notation. Table~\ref{tab:coverage-summary} summarises the key characteristics. They come from a library of canonical examples such as Needham--Schroeder public-key, Yahalom and WooLam, together with corrected variants and industry-inspired designs such as iKP, SET and EMV. These protocols define 388 distinct security goals, covering confidentiality, non-injective and injective authentication agreement, and channel guarantees. They are the set distributed with version 2025.09 of the \anbx{} compiler \cite{anbx2015}. Protocols using~\anbx{} channel notation are translated to \anb{} and saved back in \anbx{} format in order to preserve features like function type signatures.

\begin{table}[ht]
	\centering
	\small
	\begin{tabular}{l|r|r|r|r|r|r}
		\toprule
		\textbf{Category} & \textbf{Protocols} & \textbf{Goals} & \textbf{Secret} & \textbf{Auth} & \textbf{WAuth} & \textbf{Channel} \\
		\midrule
		\multicolumn{7}{c}{\textbf{Protocols}} \\
		\midrule
		Small AnB & 111 & 250 & 69 & 77 & 38 & 66 \\
		Large AnB & 4 & 19 & 7 & 2 & 9 & 1 \\
		SET family & 4 & 44 & 14 & 26 & 4 & 0 \\
		iKP family & 9 & 71 & 21 & 39 & 11 & 0 \\
		EMV & 2 & 4 & 1 & 0 & 3 & 0 \\
		\midrule
		Total & 130 & 388 & 112 & 144 & 65 & 67 \\
		\midrule
		\multicolumn{7}{c}{\textbf{Agents}} \\
		\midrule
		2 & 60 & 117 & 34 & 16 & 20 & 47 \\
		3 & 67 & 261 & 74 & 128 & 41 & 18 \\
		4 & 3 & 10 & 4 & 0 & 4 & 2 \\
		\midrule
		\multicolumn{7}{c}{\textbf{Steps}} \\
		\midrule
		1 & 20 & 29 & 3 & 2 & 8 & 16 \\
		2 & 19 & 42 & 13 & 3 & 12 & 14 \\
		3 & 22 & 49 & 15 & 15 & 5 & 14 \\
		4 & 16 & 46 & 14 & 18 & 5 & 9 \\
		5-9 & 46 & 152 & 44 & 72 & 22 & 14 \\
		10-14 & 6 & 62 & 19 & 34 & 9 & 0 \\
		15-19 & 1 & 8 & 4 & 0 & 4 & 0 \\
		\bottomrule
	\end{tabular}
	\caption{Coverage of the Test Suite}
	\label{tab:coverage-summary}
\end{table}

\begin{table}[ht]
	\centering
	\small
	\begin{tabular}{l|r|r|r|r|r|r|r|r}
		\toprule
		\textbf{Tool} & \textbf{Total} & \textbf{Attack} & \textbf{Safe} & \textbf{No Query} & \textbf{Secrecy} & \textbf{Auth} & \textbf{WAuth} & \textbf{Channel} \\
		\midrule
		OFMC 1s & 388 & 71 & 317 & 0 & 114 (24) & 144 (27) & 63 (6) & 67 (14) \\
		OFMC 2s & 245 & 86 & 159 & 0 & 68 (25) & 75 (35) & 37 (9) & 65 (17) \\
		ProVerif & 377 & 85 & 285 & 7 & 112 (23) & 142 (37) & 61 (8) & 62 (17) \\
		\bottomrule
	\end{tabular}
	\caption{Attack statistics per tool and goal category. Cells show total goals (attacks).}
	\label{tab:attack-stats}
\end{table}

\paragraph{Ground truth}
A ground-truth verdict for each goal was established using formal verification tools (Table \ref{tab:attack-stats}). ProVerif 2.05 served as the primary benchmark because it analyses an unbounded number of sessions and therefore provides the most comprehensive attack coverage~\cite{Blanchet2014}. OFMC 2022 results at two sessions (14 goals) and 1-session (4 goals)  were used only when ProVerif coverage was absent due mostly to inconclusive results, either a verdict of ``cannot be proven'' or the absence of a ProVerif query formulated to avoid loops, following a waterfall priority scheme. Of the 388 goals of the ground truth 22.9\% are vulnerable, the rest are safe.

\paragraph{Ground-truth consistency} Because the definitive verdict draws on three formal sources, we first checked how far they agree on the goals each pair jointly resolves. ProVerif and single-session OFMC agree on 348 of the 370 goals both resolve (94.1\%). Of the 22 disagreements, 19 are attacks that ProVerif detects and one-session OFMC misses, the multi-session cases that motivate ProVerif's unbounded analysis. ProVerif and two-session OFMC agree on 226 of the 231 they jointly resolve (97.8\%), and one- versus two-session OFMC differ on 20 of 245 goals, every one of them an attack visible only at two sessions. This pattern, in which more sessions expose more attacks, shows that the sources are consistent where they overlap and that their disagreements follow from the session bound rather than tool error. The pattern in turn supports the ProVerif-first waterfall. In a small number of goals, 3 against one-session and 4 against two-session OFMC, the bounded tool flags an attack that ProVerif proves absent. These cases most likely reflect differences between the ProVerif and OFMC model translations. The waterfall resolves them in ProVerif's favour, and they are natural candidates for the manual audit discussed in Section~\ref{sec:limitations}.

\paragraph{Obfuscation}

To mitigate the risk of training-data contamination, where a model might recognise a protocol by name and identifier names rather than reasoning about its structure~\cite{magar2022data,sainz2023nlp}, the protocol specifications were subjected to a \emph{type-aware identifier obfuscation} pass before being submitted to the models. We implemented a new function in the \anbx{} compiler \cite{anbx2015} that replaces every user-defined identifier such as agents, nonces, and keys with a fresh name that retains a one-letter prefix indicating the identifier’s declared type, for instance, \texttt{A} for an agent, \texttt{N} for a nonce, \texttt{K} for a symmetric key, followed by a unique numeric index. Function symbols and the Diffie–Hellman generator \texttt{g} are left untouched. Additionally, the filenames are replaced with neutral labels of the form \texttt{PROTO\_0001}. This ensures that the LLM receives a syntactically well-formed protocol whose structure is intact but whose names cannot be matched to any known protocol from its training corpus, forcing it to reason from the symbolic message sequence alone. 

While identifier renaming mitigates lexical memorisation, it does not eliminate structural memorisation: a protocol retains a characteristic message-flow signature even after renaming. The Needham–Schroeder protocol, for instance, remains recognisable by its three-message challenge–response pattern irrespective of agent names. An LLM trained on textbook analyses of this structure might reproduce a memorised verdict without genuine reasoning. We acknowledge that complete elimination of memorisation cannot be guaranteed without a synthetic-protocol benchmark, which we identify as potential future work.

\subsection{Prompt Engineering}

During pipeline development we explored several prompt variants, from a concise instruction set to the comprehensive zero-shot prompt used in the final experiment. We selected the final prompt for its consistent goal identification and high JSON compliance, enabling fully automated batch processing. Engineered to elicit structured, machine-parseable analyses without worked examples, the prompt was dispatched afresh for each protocol and comprised four components:

\begin{enumerate}
	\item \emph{Role preamble}: instructing the model to act as a symbolic security protocol verifier operating under the Dolev--Yao model.
	\item \emph{Notation reference}: defining the cryptographic operators used in \anb{}/\anbx{} with their Dolev--Yao semantics: public-key encryption and signing ($\mathsf{pk}$, $\mathsf{sk}$, $\mathsf{inv}$), symmetric encryption, hashing, tupling, and the four channel types (insecure, confidential, authenticated, secure).
	\item \emph{Goal definitions}: formally distinguishing non-injective and injective agreement, weak and strong secrecy, and channel-oriented confidentiality properties.
	\item \emph{Output schema}: a strict JSON object with five fields. \texttt{goal\_id} identifies the goal. \texttt{status} is restricted to \texttt{"attack found"} or \texttt{"no attack"}. \texttt{confidence} is an integer from 1 to 100. \texttt{justification} carries the model's reasoning, and \texttt{two\_session\_trace} is optional. The schema prohibits an ``unknown'' option to ensure every goal contributes to a binary classification.
\end{enumerate}

 The full prompt is available in Appendix~\ref{app:prompt}. It requires the model to return plain JSON without markdown wrapping and to refer to each protocol only by its anonymised identifier. The prompt contains no in-context examples of correct analysis, so the model must rely on its pre-trained knowledge of protocol analysis.

We acknowledge that our findings are conditional on the specific prompt design used. A systematic comparison of prompt variants was nonetheless beyond the scope of this study. Such a comparison would vary the amount of cryptographic guidance, use few-shot exemplars, or relax the JSON schema. We therefore caution that alternative prompt designs might produce different performance profiles, and we treat our results as indicative rather than definitive for the current generation of LLMs.

\subsection{Evaluation Metrics}
Each LLM verdict is compared with the ground-truth verdict at the level of individual security goals. We treat \texttt{"attack found"} as the positive class and \texttt{"no attack"} as the negative class, yielding the usual confusion-matrix entries:
\begin{itemize}
	\item \emph{True Positive (TP)}: LLM and ground truth both indicate an attack.
	\item \emph{True Negative (TN)}: both indicate no attack.
	\item \emph{False Positive (FP)}: LLM reports an attack where the formal tools find none.
	\item \emph{False Negative (FN)}: LLM misses an attack that the formal tools confirm.
\end{itemize}
From these counts, four summary metrics are computed per run:

\begin{equation}
\text{Accuracy} = \frac{TP + TN}{TP + TN + FP + FN}
\end{equation}

\begin{equation}
\text{Precision} = \frac{TP}{TP + FP}
\end{equation}

\begin{equation}
\text{Recall} = \frac{TP}{TP + FN}
\end{equation}

\begin{equation}
F1 = 2\cdot\frac{\text{Precision} \cdot \text{Recall}}{\text{Precision} + \text{Recall}}.
\end{equation}

Secure goals outnumber vulnerable ones in the dataset, so accuracy alone can be misleading~\cite{5128907}. In a security context, a false negative is a missed attack and carries a far higher cost than a false positive, which merely triggers a manual review. We therefore treat recall as the primary metric and report precision and F1 as complementary views~\cite{sommer2010outside}.

\paragraph{Trivial baselines}
To contextualise our results, we computed two trivial baselines directly from the class distribution of the definitive ground truth. A classifier that always predicts \texttt{no attack} achieves 77.1\% accuracy but 0\% recall, reflecting the class imbalance in the dataset, in which 22.9\% of goals carry an attack. Conversely, a classifier that always predicts \texttt{attack} achieves 100\% recall but only 22.9\% precision and 22.9\% accuracy. Table~\ref{tab:baselines} summarises these baselines.

\begin{table}[ht]
	\centering
	\small
	\begin{tabular}{l|cccc}
		\toprule
		\textbf{Baseline} & \textbf{Precision} & \textbf{Recall} & \textbf{F1} & \textbf{Accuracy} \\
		\midrule
		Always ``no attack'' & n/a & 0.0\% & 0.0\% & 77.1\% \\
		Always ``attack''    & 22.9\% & 100.0\% & 37.3\% & 22.9\% \\
		\bottomrule
	\end{tabular}
	\caption{Trivial baselines derived from the definitive ground-truth class distribution over 388 goals, of which 22.9\% are attack goals.}
	\label{tab:baselines}
\end{table}

These baselines show that accuracy alone is misleading in this imbalanced setting. They also support our decision to treat recall as the primary metric, since a missed attack carries a far higher cost than a false positive that merely triggers manual review, with precision and F1 as complementary views for evaluating the trustworthiness of attack predictions.

\paragraph{Confidence calibration}
We analyse the model’s self-reported confidence score by grouping predictions into confidence bands and plotting the proportion of correct verdicts in each band. A well-calibrated model should show accuracy rising monotonically with confidence. A flat or erratic relationship indicates unreliable self-assessment.

\paragraph{Inter-run consistency}
For each model configuration and each security goal, we record the verdict across all runs. A goal is \emph{unanimous} when every run agrees and \emph{split} when verdicts differ, and we take the majority vote as the aggregate verdict. This analysis exposes the protocols on which the model’s reasoning is unstable, and it measures how much a simple ensemble can improve overall accuracy~\cite{Trad2024}.
\\
\paragraph{Statistical stability}
Because goals are associated with protocols, treating individual goals as independent observations would constitute pseudo-replication~\cite{Hurlbert1984} and produce artificially narrow confidence intervals. To obtain honest uncertainty estimates, we compute 95\% confidence intervals via a cluster bootstrap~\cite{Field2007} that resamples protocols (with replacement) rather than individual goals. Each bootstrap iteration draws 130 protocols from the original set, preserves all goals within each sampled protocol, and recomputes precision, recall, F1, and accuracy. The 2.5th and 97.5th percentiles of the resulting distributions form the 95\% confidence interval. This approach correctly accounts for the natural correlation among goals within the same protocol and reflects the variability expected when applying the method to a different set of protocols.

\section{Implementation}
\label{sec:implementation}

The experimental workflow was realised as a modular Python pipeline whose core is the batch-processing script.\footnote{The source code, along with all experimental artefacts, is available at \url{https://github.com/arquam-hub/protocol-checker-}.} The pipeline automates the full cycle from protocol loading to metric computation, ensuring reproducibility and eliminating manual transcription errors.

\subsection{System Architecture}
\begin{figure}[t]
		\centering
		\small
		\begin{tikzpicture}[
			node distance = 0.8cm and 0.8cm,
			box/.style = {
				rectangle, rounded corners, draw=black, thick,
				fill=blue!5, text width=3.5cm, align=center,
				minimum height=1.2cm, font=\sffamily\scriptsize
			},
			widebox/.style = {box, text width=5cm},
			subbox/.style = {box, fill=green!5, text width=2.4cm, minimum height=0.9cm,
				font=\sffamily\footnotesize},
			evalbox/.style = {box, fill=orange!10, text width=3.5cm},
			arrow/.style = {->, >=Stealth, thick},
			label/.style = {font=\sffamily\scriptsize\itshape}
			]
			
			\node[box] (load) {\textbf{1. Protocol Loading \& Parsing}\\ 
				Read \texttt{.AnBx} files,\\ extract body and goals,\\ strip comments};
			\node[box, below=of load] (anon) {\textbf{2. Obfuscation}\\ 
				Replace protocol names\\ with neutral IDs\\ (\texttt{PROTO\_0001})};
			\node[box, below=of anon] (prompt) {\textbf{3. Prompt Construction}\\ 
				Combine obfuscated body\\ with fixed system message\\ and JSON schema};
			\node[widebox, below=of prompt] (dispatch) {\textbf{4. API Dispatch}\\
				Submit prompt concurrently to\\ OpenAI and DeepSeek APIs\\
				via \texttt{ThreadPoolExecutor}};
			\node[subbox, below left=0.8cm and -1.4cm of dispatch] (openai) {OpenAI\\ GPT-5.4};
			\node[subbox, below right=0.8cm and -1.4cm of dispatch] (deepseek) {DeepSeek\\ v4 Flash};
			\draw[arrow, dashed, gray] (dispatch.south) -- ++(0,-0.4) -| (openai.north);
			\draw[arrow, dashed, gray] (dispatch.south) -- ++(0,-0.4) -| (deepseek.north);
			
			\node[box, below=2.4cm of dispatch] (parse) {\textbf{5. Response Parsing \& Export}\\ 
				Validate JSON, normalise,\\ write verdicts to CSV\\ (one row per goal)};
			\draw[arrow] (openai.south) -- ++(0,-0.3) -| ([xshift=-1cm]parse.north);
			\draw[arrow] (deepseek.south) -- ++(0,-0.3) -| ([xshift=1cm]parse.north);
			
			\node[evalbox, right=2.2cm of parse] (eval) {\textbf{Evaluation Script}\\ 
				computes precision,\\
				recall, F1, confidence\\
				calibration, per-goal\\ breakdowns};
			\draw[arrow] (parse.east) -- node[above, align=center, font=\footnotesize\sffamily] {CSV output} (eval.west);
			
			\draw[arrow] (load) -- (anon);
			\draw[arrow] (anon) -- (prompt);
			\draw[arrow] (prompt) -- (dispatch);
			
			\node[box, fill=gray!10, above=0.8cm of eval] (truth) {ProVerif / OFMC\\ Ground Truth};
			\draw[arrow] (truth.south) -- (eval.north);
			
		\end{tikzpicture}
		\caption{Workflow of the protocol-analysis pipeline}
		\label{fig:system-workflow}
	\end{figure}
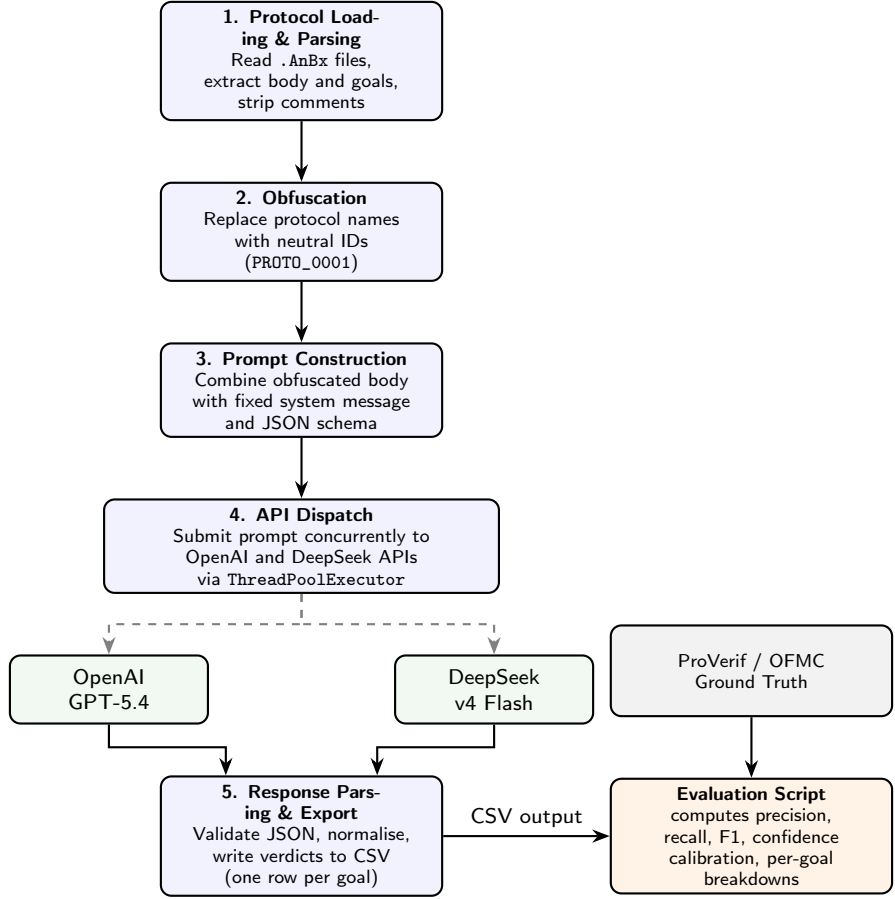

The system operates as illustrated in Figure~\ref{fig:system-workflow}:
\begin{enumerate}
	\item \emph{Protocol loading and parsing.} The pipeline reads \anb{}/\anbx{} files from an input directory and the parser strips all comments and extracts the protocol body, which holds the message exchange, together with the list of security goals.
	\item \emph{Protocol name obfuscation.} The pipeline replaces each protocol name with a neutral label of the form \texttt{PROTO\_0001}, even when the name is already anonymised. The protocol specifications are already obfuscated as detailed in Section \ref{subsec:obfuscation}.
	\item \emph{Prompt construction.} The pipeline builds a structured prompt, reproduced in Appendix~\ref{app:prompt}, by combining the obfuscated protocol body with a fixed system message that defines the cryptographic notation, the Dolev--Yao semantics, the security goal types, and the required JSON output schema.
	\item \emph{API dispatch.} The pipeline submits the prompt concurrently to the OpenAI and DeepSeek APIs using Python's \texttt{ThreadPoolExecutor}, and maintains a separate model-priority list for each provider and mode combination. A per-request timeout guards against unresponsive endpoints, and the pipeline retries failed calls using the next available model in the priority list.
	\item \emph{Response parsing and export.} The pipeline validates and normalises the JSON body of each API response, then writes it to a CSV file with one row per model per protocol goal. Each row records the goal identifier, the binary verdict, the self-reported confidence score, the textual justification, and metadata such as the timestamp and token usage. The verdict field takes the value \texttt{attack\_found} or \texttt{no\_attack}.
\end{enumerate}

A final stage of the same pipeline computes the evaluation metrics. It reads a comparison table that pairs each model's verdict with the ground-truth label for every protocol goal, then sorts each case into a true or false positive or negative. From these counts it derives the precision, recall, F1 score, and accuracy for each model against each truth source. The stage writes two files: one row per case, and a summary of the counts and derived scores reported in Section~\ref{sec:evaluation}.

\subsection{Reasoning Mode Configuration}
\label{sec:reasoning-impl}
The two providers expose reasoning capabilities through different API interfaces, requiring distinct handling.

\paragraph{OpenAI}
Both modes send the same model, \texttt{gpt-5.4}, to the Responses API with the same system and user messages. They differ only in the reasoning-effort parameter: \texttt{reasoning=\{"effort": "none"\}} in chat mode and \texttt{reasoning=\{"effort": "high"\}} in reasoning mode. The latter instructs the model to allocate additional computational budget to its internal chain-of-thought process. Neither request carries a temperature, so the two arms match on sampling as well as on model. The system maintains a set of model identifiers designated for the Responses API (\texttt{RESPONSES\_API\_MODELS}) and routes calls accordingly. The lower-priority fallback models sit on the Chat Completions endpoint with a temperature of $0.1$, and a result produced by one of them is labelled as such rather than recorded as the arm's controlled condition. A provider rejection of the reasoning-effort setting ends that request as a failure instead of silently advancing to a model that would have reasoned.

\paragraph{DeepSeek}
DeepSeek exposes an OpenAI-compatible REST interface, so the pipeline sends both chat and reasoning requests through the same OpenAI client library, pointed at the DeepSeek base URL. The \texttt{deepseek-reasoner} endpoint maps to the DeepSeek v4 Flash model, which rejects several parameters the chat models accept. For that endpoint the pipeline strips those parameters, such as \texttt{temperature}, from the request payload before sending it, which avoids the API errors documented by DeepSeek~\cite{deepseekReasoningDocs}. As with OpenAI, the pipeline keeps a prioritised fallback list of model identifiers for each provider and mode combination, so it can move to the next model when an endpoint fails.

\subsection{Evaluation and Output Pipeline}
\label{sec:eval-pipeline}
The raw verdicts produced by the LLMs are evaluated by a set of dedicated routines in a Python program, which implement the following logic:
\begin{itemize}
	\item \emph{Ground-truth matching.} Each LLM output row is paired with the corresponding ground-truth entry using the normalised protocol identifier and goal. A waterfall priority scheme (ProVerif $\to$ OFMC 2-session $\to$ OFMC 1-session) selects the definitive ground-truth verdict per goal; a higher-priority source is used only when it yields a usable \textsc{attack}/\textsc{no\_attack} verdict, so a lower-priority source resolves goals the preferred source leaves uncovered.
	\item \emph{Confusion matrix computation.} For each model, the pipeline counts true positives, false positives, true negatives, and false negatives, both against each individual ground-truth source and against the definitive verdict. We then derive accuracy, precision, recall, and F1-score from shared formulas.
	\item \emph{Goal-type disaggregation.} The same metrics are computed separately for each security-goal category (confidentiality, secrecy, injective agreement, non-injective agreement, authenticated channel, and secure channel), with goals categorised from their \anbx{} statements; the goal verb separates injective agreement (plain \texttt{authenticates}) from non-injective agreement (\texttt{weakly authenticates}), and bare message-transmission goals that match no category are excluded.
	\item \emph{Confidence calibration.} Predictions carrying a usable confidence score are grouped by self-reported confidence interval (90--100, 80--89, \dots), and the fraction of correct verdicts, which is judged against the definitive verdict, is computed within each interval, emitted both as a CSV table and as a rendered reliability diagram.
	\item \emph{Inter-run consistency.} For configurations with multiple runs, the pipeline aggregates the per-run comparison outputs and records per-goal verdict agreement, the modal verdict, the agreement fraction, and a unanimity flag. It then computes majority-vote metrics and excludes tied goals from the aggregate confusion matrix.
\end{itemize}
Each analysis is invoked independently and writes its results to CSV for manual inspection and downstream plotting.

\section{Evaluation}
\label{sec:evaluation}

\subsection{Experimental Setup}
\label{sec:exp-setup}
All experiments were conducted using the pipeline described in Section~\ref{sec:implementation}. Every configuration is a remote API call, so the client operating system has no bearing on the verdicts: the pipeline only composes the request, parses the returned JSON, and writes it to CSV. The runs reported here were launched from more than one workstation over the course of the study, and we record that only for completeness. The pipeline runs on any reasonably modern machine with Python v3.12. API calls were scheduled during off-peak hours to minimise latency and rate-limiting. The chat models were run on Windows~11 and the reasoning models on macOS~26.5. This distributed the workload between authors and does not affect the results, since API-based inference is independent of the operating system.

\paragraph{Reproducibility}
The four configurations are \texttt{gpt-5.4} at reasoning effort \texttt{none} for chat and \texttt{high} for reasoning, both through the OpenAI Responses API, and \texttt{deepseek-v4-flash} without and with thinking through the DeepSeek API. Neither OpenAI request sends a temperature. The DeepSeek reasoning endpoint rejects several parameters the chat endpoint accepts, and the pipeline strips those before sending. The reasoning runs were executed on 17 June 2026, the DeepSeek chat runs on 21 July 2026, and the GPT chat runs on 26 August 2026, three independent runs each over all 130 protocols. Each provider is configured with a prioritised fallback list, and a request that fails transiently retries before the pipeline advances to the next candidate. Every result row records the model and mode that produced it in a \texttt{model\_variant} column, so a fallback is visible in the data rather than hidden in a log. No fallback model was invoked in any run. The chat arm also reports zero reasoning tokens across all three runs, which is consistent with the provider having honoured the request to disable reasoning rather than merely accepted the parameter.

\emph{Reasoning-model runs} Each reasoning model was evaluated over three independent runs on the full dataset of 130 protocols and 388 security goals, matching the chat-model protocol.

Table~\ref{tab:api-pricing} summarises the per-token prices at the time of the study~\cite{OpenAI2026APIPricing,DeepSeek2026ModelsPricing}.

\begin{table}[t]
	\centering
	\caption{API pricing per million tokens (chat and reasoning modes). Note: the pricing for DeepSeek v4-flash is from June 2026}
	\label{tab:api-pricing}
	\begin{tabular}{llrrr}
		\hline
		\emph{Provider} & \emph{Model} & \emph{Input} & \emph{Output} & \emph{Cached Input} \\
		\hline
		OpenAI & GPT-5.4 (both modes) & \$2.50 & \$15.00 & \$0.25 \\
		DeepSeek & v4 Flash (both modes) & \$0.14 & \$0.28 & \$0.0028 \\
		\hline
	\end{tabular}
\end{table}

\subsection{Chat Model Results}
\label{sec:chat-results}
Tables~\ref{tab:chat-confusion} and~\ref{tab:chat-metrics} present the per-run classification counts and derived metrics for both chat models. The results reveal a consistent pattern. Both models achieve moderate recall but poor precision. GPT chat attains an average recall of 72.7\%, correctly flagging most true attacks, but its precision averages only 27.3\%, meaning that fewer than one in three attack claims is justified. DeepSeek chat is close on both counts, at 69.3\% recall and 27.2\% precision. The two arms are statistically indistinguishable. On the 388 paired goals of the first run, a McNemar exact test on per-goal correctness (comparing, for each goal, whether GPT and DeepSeek were correct or incorrect) gives $p=0.72$, with 61 goals correct for GPT alone against 66 for DeepSeek alone. This high p-value supports the claim that the two chat configurations do not differ in accuracy. The large difference in precision and recall that distinguishes the providers in reasoning mode is absent here.

The average accuracy of both models sits near 49 to 50\%, which is below the 77.1\% of the majority-class baseline in Table~\ref{tab:baselines}. That gap follows from their false-positive volume rather than showing that they are uninformative, since each roughly doubles the trivial precision of 22.9\% while recovering around 70\% of the attacks. This gap illustratse why Section~\ref{sec:methodology} nominates recall, not accuracy, as the primary metric for this task.

\begin{table}[t]
	\centering
	\caption{Per-run classification counts for chat models (388 goals)}
	\label{tab:chat-confusion}
	\small
	\begin{tabular}{llrrrr}
		\hline
		\emph{Model} & \emph{Run} & \emph{TP} & \emph{FP} & \emph{TN} & \emph{FN} \\
		\hline
		\multirow{3}{*}{DeepSeek Chat} & 1 & 63 & 174 & 125 & 26 \\
		& 2 & 59 & 163 & 136 & 30 \\
		& 3 & 63 & 158 & 141 & 26 \\
		\hline
		\multirow{3}{*}{GPT Chat}      & 1 & 65 & 171 & 128 & 24 \\
		& 2 & 66 & 179 & 120 & 23 \\
		& 3 & 63 & 167 & 132 & 26 \\
		\hline
	\end{tabular}
\end{table}

\begin{table}[t]
	\centering
	\caption{Per-run metrics for chat models.}
	\label{tab:chat-metrics}
	\small
	\begin{tabular}{llrrrr}
		\hline
		\emph{Model} & \emph{Run} & \emph{Precision} & \emph{Recall} & \emph{F1} & \emph{Accuracy} \\
		\hline
		\multirow{3}{*}{DeepSeek Chat} & 1 & 26.6\% & 70.8\% & 38.7\% & 48.5\% \\
		& 2 & 26.6\% & 66.3\% & 37.9\% & 50.3\% \\
		& 3 & 28.5\% & 70.8\% & 40.6\% & 52.6\% \\
		\hline
		\multirow{3}{*}{GPT Chat}      & 1 & 27.5\% & 73.0\% & 40.0\% & 49.7\% \\
		& 2 & 26.9\% & 74.2\% & 39.5\% & 47.9\% \\
		& 3 & 27.4\% & 70.8\% & 39.5\% & 50.3\% \\
		\hline
	\end{tabular}
\end{table}

Neither model drifts in a consistent direction across the three runs. GPT’s false negatives move 24, 23, 26 and its false positives 171, 179, 167, so the run-to-run spread is real but non-monotone. DeepSeek’s false positives ease from 174 to 158 while its false negatives hold near 26 to 30 and recall stays close to 70\%. With only three runs we cannot distinguish a gradual shift in a model’s decision tendency from the ordinary non-determinism of LLM generation, and we make no claim about the mechanism. The variation establishes its own magnitude only, which Section~\ref{sec:consistency} quantifies per goal.

\subsection{Reasoning Model Results}
\label{sec:reasoning-results}

Both reasoning models were evaluated over three independent runs on the full dataset of 130 protocols and 388 security goals. GPT-5.4 was accessed through the OpenAI API, and DeepSeek~v4~Flash through DeepSeek API. Verdicts are scored against three formal ground-truth sources: OFMC restricted to a single session, OFMC at two sessions, and ProVerif. OFMC at one session resolves all 388 goals. ProVerif resolves 370, excluding 18 goals it cannot prove. The two-session source resolves only 245, with the rest failing to terminate within the bounded state-space.
Tables~\ref{tab:reasoning-full-confusion} and~\ref{tab:reasoning-full} report, respectively, the per-run classification counts against the ProVerif benchmark and the run-averaged metrics against each source.

Against ProVerif, the primary benchmark, GPT reasoning attains an average precision of 66.5\% and recall of 54.5\% (F1 59.9\%), while DeepSeek attains 45.4\% precision and 57.3\% recall (F1 50.6\%). Relative to chat mode, both models trade a large share of their false positives for higher precision: against the same ProVerif benchmark GPT's precision rises from 26.8\% to 66.5\% and DeepSeek's from 27.2\% to 45.4\%. Both reasoning models detect a slight majority of the true attacks, 54 to 57\%. Unlike the chat models, both achieve higher precision at the cost of reduced recall, 54.5\% for GPT and 57.3\% for DeepSeek. Both contrasts isolate the effect of reasoning itself, because each provider holds its model fixed across the two modes and varies only whether it reasons. For GPT the two arms are the same \texttt{gpt-5.4} endpoint at reasoning effort \texttt{none} and \texttt{high}. They consumed an identical 441{,}972 prompt tokens over the three runs, so the same prompt reached the same model in both, and they differ only in what the model generated in response. For DeepSeek the two arms are v4~Flash run without and with its chain-of-thought. Accuracy is high for GPT (83.2\%) and moderate for DeepSeek (74.4\%), because DeepSeek produces far more false positives, which results in 58.3 per run on average, against 23.3 for GPT.

Table~\ref{tab:reasoning-full} shows that precision varies widely with the ground truth: it is highest against OFMC 2-session (GPT 74.7\%, DeepSeek 61.3\%), where a second session confirms many of the goals the models flag, and lowest against OFMC 1-session (GPT 54.3\%, DeepSeek 37.1\%), the most conservative benchmark. Recall, by contrast, stays close to constant across sources (GPT 53--56\%, DeepSeek 55--58\%). The attacks the models miss, regardless of which tool defines the ground truth, represent a structural blind spot rather than an artefact of benchmark choice. On recall, the metric we prioritise in Section~\ref{sec:methodology}, the two models are comparable. GPT leads DeepSeek on precision, F1, and accuracy against every source.

To quantify \emph{stability}, we compute 95\% confidence intervals via a protocol-level cluster bootstrap (resampling protocols with replacement, preserving each protocol's goal set, and pooling the three runs) so the intervals align with the run-averaged point estimates. For the definitive verdict, GPT's precision is 64.8\% (95\% CI:~[49.1, 78.1]) and recall is 53.2\% (95\% CI:~[39.3, 66.3]). DeepSeek's precision is 44.4\% (95\% CI:~[30.7, 58.5]) and recall is 55.1\% (95\% CI:~[41.6, 67.4]). These marginal intervals overlap, but overlap between the intervals does not constitute a test of the difference. Because both models are scored on the same resampled protocols, we bootstrap the difference itself: GPT's precision advantage is $+20.4$ percentage points with a 95\% CI:~[9.1, 31.4], which excludes zero, whereas the recall difference is $-1.9$ points with a 95\% CI:~[-8.9, 4.8], which does not. GPT is therefore more precise than DeepSeek at the protocol level, with no detectable difference in recall. The width of these intervals reflects the diversity of the 130 protocols in our benchmark: performance varies substantially across protocol families, and this variation is properly reflected in the protocol-level bootstrap.
 
The \emph{definitive} verdict consolidates the three sources under the waterfall priority of Section~\ref{sec:methodology}, resolving all 388 goals. Because ProVerif supplies the verdict wherever it terminates, the definitive figures track the ProVerif row closely (GPT 64.8\% precision, 53.2\% recall, 82.6\% accuracy, and DeepSeek 44.4\%, 55.1\%, 73.9\%), with the 18 ProVerif-uncovered goals filled by OFMC. We treat this consolidated verdict as the reference benchmark in the analyses that follow.

\begin{table}[t]
	\centering
	\caption{Per-run classification counts for reasoning models on the full dataset (ProVerif ground truth, 370 goals)}
	\label{tab:reasoning-full-confusion}
	\small
	\begin{tabular}{llrrrr}
		\hline
		\emph{Model} & \emph{Run} & \emph{TP} & \emph{FP} & \emph{TN} & \emph{FN} \\
		\hline
		\multirow{3}{*}{GPT-5.4} & 1 & 47 & 23 & 262 & 38 \\
		& 2 & 43 & 24 & 261 & 42 \\
		& 3 & 49 & 23 & 262 & 36 \\
		\hline
		\multirow{3}{*}{DeepSeek v4 Flash} & 1 & 54 & 60 & 225 & 31 \\
		& 2 & 47 & 56 & 229 & 38 \\
		& 3 & 45 & 59 & 226 & 40 \\
		\hline
	\end{tabular}
\end{table}

\begin{table}[t]
	\centering
	\caption{Reasoning models on the full dataset, averaged over three runs, against each ground-truth source and the definitive (waterfall) verdict. 95\% confidence intervals for the definitive verdict are reported in the text.}
	\label{tab:reasoning-full}
	\small
	\begin{tabular}{llrrrrr}
		\hline
		\emph{Model} & \emph{Ground truth} & \emph{Goals} & \emph{Prec.} & \emph{Rec.} & \emph{F1} & \emph{Acc.} \\
		\hline
		\multirow{4}{*}{GPT-5.4}
		& Definitive     & 388 & 64.8\% & 53.2\% & 58.4\% & 82.6\% \\
		& OFMC 1-session & 388 & 54.3\% & 55.9\% & 55.1\% & 83.3\% \\
		& OFMC 2-session & 245 & 74.7\% & 53.9\% & 62.6\% & 77.4\% \\
		& ProVerif       & 370 & 66.5\% & 54.5\% & 59.9\% & 83.2\% \\
		\hline
		\multirow{4}{*}{DeepSeek v4 Flash}
		& Definitive     & 388 & 44.4\% & 55.1\% & 49.1\% & 73.9\% \\
		& OFMC 1-session & 388 & 37.1\% & 57.7\% & 45.1\% & 74.4\% \\
		& OFMC 2-session & 245 & 61.3\% & 56.2\% & 58.7\% & 72.2\% \\
		& ProVerif       & 370 & 45.4\% & 57.3\% & 50.6\% & 74.4\% \\
		\hline
	\end{tabular}
\end{table}

\subsection{Performance by Goal Type}
\label{sec:goal-type}
Disaggregating the results by security goal category reveals substantial variation in model capability illustrated in Tables~\ref{tab:goal-type-chat} and~\ref{tab:goal-type-reasoning}. Goal categories are derived from the \anbx{} goal expression, which includes a small number of bare message-transmission goals with the form $A \to B : M$ to fall under none of the six categories. These goals are omitted from this breakdown. Confidentiality goals are classified most reliably by all models. Averaged over the three reasoning runs, GPT achieves an F1 of 95.7\% on confidentiality, with perfect recall and a precision of 91.7\%.

Secrecy and authentication goals present a more mixed picture. In chat mode, both models over-flag secrecy goals while still missing some genuine secrecy attacks. Authenticated channels are over-flagged hardest. Both chat models predict an attack on every authenticated-channel goal, attaining 100\% recall at roughly 12 to 15\% precision. This over-prediction is not demonstrated in the reasoning mode. The category contains very few attack goals in the definitive set, so its rates are noisy. DeepSeek catches the single case it faces while GPT recovers only about one in three across runs, but the indiscriminate flagging is gone. The shift from over-prediction to erratic, sparse handling suggests that the models lack both a principled understanding of channel guarantees and a consistent decision policy.

Injective agreement remains the hardest well-populated category, and here the sample is large enough to draw a conclusion that the 388-goal set contains 39 injective-agreement attacks spread over 27 protocols. Averaged over the three runs, GPT recovers only about 15 of them, a recall of 38.5\% (95\% CI:~[22.4, 55.3]), and DeepSeek reaches 40.2\% (95\% CI:~[24.1, 57.7]). Both intervals include 50\%, so the data do not rule out a true detection rate slightly above half. Neither upper limit reaches 58\%, however, so at best the models recover a small majority of injective-agreement attacks. Secrecy, the other well-populated category with 24 attack goals over 16 protocols, is resolved no better. GPT recovers 65.3\% (95\% CI:~[38.3, 89.3]) and DeepSeek 62.5\% (95\% CI:~[36.2, 84.4]), intervals that also include 50\%. GPT's high precision on injective agreement (82.8\%) reflects that the few attacks it does flag are usually real rather than broad coverage. The remaining authentication categories rest on far fewer attack goals, such as non-injective agreement on 9, authenticated channel on a single attack goal, and secure channel on 5. Their point rates are therefore descriptive only, with non-injective recall at 25.9\% for GPT (95\% CI:~[0.0, 52.4]) and 29.6\% for DeepSeek (95\% CI:~[5.6, 61.1]), and their wide intervals preclude strong per-category claims. Read together, the well-powered injective-agreement result and the uniformly low authentication recall still indicate that the models do not meaningfully differentiate the two agreement strengths.

The chat models recover the same categories more often, but the intervals show that the gain buys little discrimination. On injective agreement, GPT chat reaches a recall of 79.5\% with a protocol-cluster 95\% CI:~[69.7, 88.9] and DeepSeek chat 56.4\% (95\% CI:~[42.5, 70.6]), against precisions of 32.4\% and 28.7\%. On non-injective agreement, the recalls are 55.6\% for GPT (95\% CI:~[20.0, 100.0]) and 74.1\% for DeepSeek (95\% CI:~[46.7, 95.2]), at precisions of 12.5\% and 14.9\%. On secrecy the recalls are 69.4\% for GPT (95\% CI:~[51.7, 87.0]) and 72.2\% for DeepSeek (95\% CI:~[53.9, 87.0]). Interval width varies enough across categories that the rates do not all carry equal weight. The narrowest belongs to DeepSeek on confidentiality (95\% CI:~[81.8, 100.0]), which excludes every value below 81.8\%, and GPT's injective-agreement interval is comparably tight. The widest is GPT on non-injective agreement (95\% CI:~[20.0, 100.0]), which spans almost the whole range and supports no per-category claim. The five secure-channel attacks are similarly thin. The contrast that survives the uncertainty is the one visible in the two tables. Chat models buy their agreement-goal recall with precision below 33\%, whereas reasoning models hold much higher precision and still recover well under half of the same attacks.

\begin{table}[t]
	\centering
	\caption{Chat model performance by goal type, averaged over three runs against the definitive verdict. \emph{Attacks} is the number of attack goals in the category, shared by both models because the truth is fixed, and identical to Table~\ref{tab:goal-type-reasoning}. Protocol-cluster bootstrap 95\% confidence intervals for the recall figures are reported in the text. Categories with very few attack goals, namely authenticated channel, secure channel, and non-injective agreement, are descriptive only.}
	\label{tab:goal-type-chat}
	\small
	\begin{tabular}{llrrrr}
		\hline
		\emph{Goal Type} & \emph{Model} & \emph{Attacks} & \emph{Avg Prec.} & \emph{Avg Recall} & \emph{Avg F1} \\
		\hline
		Confidentiality    & DeepSeek & 11 & 48.4\% & 93.9\% & 63.8\% \\
		Confidentiality    & GPT      & 11 & 57.9\% & 78.8\% & 66.4\% \\
		Secrecy            & DeepSeek & 24 & 30.3\% & 72.2\% & 42.7\% \\
		Secrecy            & GPT      & 24 & 27.1\% & 69.4\% & 38.9\% \\
		Injective Agreement  & DeepSeek & 39 & 28.7\% & 56.4\% & 38.0\% \\
		Injective Agreement  & GPT      & 39 & 32.4\% & 79.5\% & 46.0\% \\
		Non-inj. Agreement   & DeepSeek & 9 & 14.9\% & 74.1\% & 24.8\% \\
		Non-inj. Agreement   & GPT      & 9 & 12.5\% & 55.6\% & 20.4\% \\
		Auth.\ Channel     & DeepSeek & 1 & 14.5\% & 100.0\% & 25.3\% \\
		Auth.\ Channel     & GPT      & 1 & 11.6\% & 100.0\% & 20.7\% \\
		Secure Channel     & DeepSeek & 5 & 35.1\% & 86.7\% & 49.9\% \\
		Secure Channel     & GPT      & 5 & 31.0\% & 46.7\% & 37.2\% \\
		\hline
	\end{tabular}
\end{table}

\begin{table}[t]
	\centering
	\caption{Reasoning model performance by goal type, averaged over three runs against the definitive verdict. \emph{Attacks} is the number of attack goals in the category, shared by both models because the truth is fixed. Protocol-cluster bootstrap 95\% confidence intervals for the agreement categories are reported in the text. Categories with very few attack goals, namely authenticated channel, secure channel, and non-injective agreement, are descriptive only.}
	\label{tab:goal-type-reasoning}
	\small
	\begin{tabular}{llrrrr}
		\hline
		\emph{Goal Type} & \emph{Model} & \emph{Attacks} & \emph{Avg Prec.} & \emph{Avg Recall} & \emph{Avg F1} \\
		\hline
		Confidentiality    & DeepSeek & 11 & 87.0\% & 97.0\% & 91.5\% \\
		Confidentiality    & GPT      & 11 & 91.7\% & 100.0\% & 95.7\% \\
		Secrecy            & DeepSeek & 24 & 47.1\% & 62.5\% & 53.6\% \\
		Secrecy            & GPT      & 24 & 54.1\% & 65.3\% & 59.1\% \\
		Injective Agreement  & DeepSeek & 39 & 43.0\% & 40.2\% & 41.5\% \\
		Injective Agreement  & GPT      & 39 & 82.8\% & 38.5\% & 52.3\% \\
		Non-inj. Agreement   & DeepSeek & 9 & 33.3\% & 29.6\% & 30.6\% \\
		Non-inj. Agreement   & GPT      & 9 & 93.3\% & 25.9\% & 37.8\% \\
		Auth.\ Channel     & DeepSeek & 1 & 26.1\% & 100.0\% & 41.1\% \\
		Auth.\ Channel     & GPT      & 1 & 16.7\% & 33.3\% & 22.2\% \\
		Secure Channel     & DeepSeek & 5 & 46.6\% & 80.0\% & 58.1\% \\
		Secure Channel     & GPT      & 5 & 68.3\% & 60.0\% & 63.7\% \\
		\hline
	\end{tabular}
\end{table}

\subsection{False-Negative and False-Positive Analysis}
\label{sec:fn-fp}
\paragraph{Persistent false negatives}
Several protocols were missed by all models across every run. In the chat experiments, four goals in four protocols were missed by both models in all three runs: \texttt{NSPK}, \texttt{WooLamMutual}, \texttt{Secret\_Goal\_Pair}, and the original version of the e-commerce protocol \emph{iKP}. Each model also carries its own larger blind spot. GPT chat misses 12 goals across 10 protocols in every run, adding \texttt{Bullet\_Channel}, \texttt{ForwardAny}, \texttt{ISO4pass}, \texttt{TMN}, and the key exchanges \texttt{KeyEx2} and \texttt{KeyEx5b}. The reasoning models, despite their higher precision, inherited most of these blind spots. In the full-dataset run, the two reasoning models shared 39 common false negatives, including the same set of missed protocols. These persistently missed cases involve injective-agreement violations and secrecy breaches in protocols that rely on a trusted third party or on multi-session interleaving.

\paragraph{False positives and ground-truth sensitivity}
The high false-positive rate observed in chat mode is partly attributable to the use of ProVerif as the primary benchmark. Re-evaluating the same LLM verdicts against OFMC 1-session raises the apparent false-positive count further, since OFMC 1-session detects fewer attacks by missing those on parallel sessions. However, the \emph{missed attacks} (false negatives) do not disappear under any ground-truth source. Both reasoning models fail to detect the majority of multi-session attacks that ProVerif identifies but OFMC 1-session does not. This confirms that the models’ failures are not artefacts of a particular ground-truth choice but reflect an inability to reason about concurrent session interactions.

\subsection{Inter-Run Consistency}
\label{sec:consistency}
The three-run experiments allow an assessment of verdict stability. In the chat-model runs, GPT’s false negatives move 24, 23, 26 and its false positives 171, 179, 167, whereas DeepSeek’s false negatives hold near 26 to 30 while its false positives ease from 174 to 158. Across the three full-dataset reasoning runs, GPT stays stable at 64 to 68\% precision, with 23 to 24 false positives against ProVerif, whereas DeepSeek's precision drifts downward from run to run, from 47\% to 43\%, on a larger false-positive count.

Quantifying this per goal gives a clean ordering across all four configurations. GPT reasoning returns an identical verdict across all three runs on 89.7\% of the 388 goals, DeepSeek reasoning on 74.0\%, GPT chat on 70.1\%, and DeepSeek chat on 61.6\%. Stability therefore tracks both axes at once. Within each provider, reasoning is more stable than chat, and within each mode, GPT is more stable than DeepSeek. The least precise configurations are also the least stable ones. Disagreement across runs is itself a useful signal. Between 10\% of goals for GPT reasoning and 38\% for DeepSeek chat change verdict between runs, and these are the goals on which the model's reasoning is least settled. They may warrant particular attention in a screening workflow.

Aggregating the three runs by majority vote yields a small gain in reasoning mode for the less stable configuration. Against the definitive verdict, DeepSeek reasoning's accuracy rises from a per-run average of 73.9\% to 75.5\% and its precision from 44.4\% to 47.1\%. GPT reasoning, already stable, is essentially unchanged, moving from 82.6\% to 83.0\% accuracy and 64.8\% to 67.2\% precision. In chat mode the same aggregation buys nothing. GPT chat moves from 49.3\% to 49.5\% accuracy and 27.3\% to 27.0\% precision, and DeepSeek chat from 50.4\% to 51.6\% and 27.2\% to 28.2\%. Ensemble aggregation therefore helps where a model is otherwise sound but noisy, and does not rescue one whose errors are systematic. Averaging three runs cannot remove a false positive that all three runs agree on.

\subsection{Confidence Calibration}
\label{sec:confidence}
\begin{figure}[tb]
	\centering
	\pgfplotstableread[col sep=comma]{anc/calibration_pooled_chat.csv}\calibchat
	\pgfplotstableread[col sep=comma]{anc/calibration_pooled_gpt.csv}\calibgptreason
	\pgfplotstableread[col sep=comma]{anc/calibration_pooled_deepseek.csv}\calibdsreason
	\begin{tikzpicture}
		\begin{axis}[
			xlabel = {Reported confidence (\%), bin midpoint},
			ylabel = {Fraction of verdicts correct},
			xmin = 55, xmax = 100,
			ymin = 0, ymax = 1.08,
			clip mode = individual,
			width = \textwidth,
			height = 8.5cm,
			grid = both,
			legend cell align = left,
			xtick = {60,70,80,90,100},
			ytick = {0,0.2,0.4,0.6,0.8,1.0},
			tick label style = {font=},
			label style = {font=},
			legend columns = 3,
				legend style = {at={(0.5,-0.30)}, anchor=north, font=\footnotesize}
			]
				\addplot[
				domain = 55:100,
				dashed, thick, black
				] {x/100};
				\addlegendentry{Perfect calibration}

				\addplot[
				mark = o,
				dashed, thick,
				blue,
				mark options = {solid},
				point meta = explicit,
				nodes near coords = {$n$=\pgfmathprintnumber[precision=0,1000 sep={,}]{\pgfplotspointmeta}},
				every node near coord/.append style = {font=\small, text=blue, anchor=north east, yshift=-3pt, xshift=1pt},
				skip coords between index = {4}{7} %
				] table[
				x expr = {\thisrow{lower}+5},
				y = fraction_correct,
				meta = count
				] {\calibchat};
				\addlegendentry{GPT (chat), $N$=1{,}164}

				\addplot[
				mark = square,
				dashed, thick,
				red,
				mark options = {solid},
				point meta = explicit,
				nodes near coords = {$n$=\pgfmathprintnumber[precision=0,1000 sep={,}]{\pgfplotspointmeta}},
				every node near coord/.append style = {font=\small, text=red, anchor=south, yshift=2pt},
				skip coords between index = {0}{4} %
				] table[
				x expr = {\thisrow{lower}+5},
				y = fraction_correct,
				meta = count
				] {\calibchat};
				\addlegendentry{DeepSeek (chat), $N$=1{,}126}

				\addplot[
				mark = *,
				thick,
				blue,
				mark options = {solid},
				point meta = explicit,
				nodes near coords = {$n$=\pgfmathprintnumber[precision=0,1000 sep={,}]{\pgfplotspointmeta}},
				every node near coord/.append style = {font=\small, text=blue, anchor=north, yshift=-2pt}
				] table[
				x expr = {\thisrow{lower}+5},
				y = fraction_correct,
				meta = count
				] {\calibgptreason};
				\addlegendentry{GPT (reasoning), $N$=1{,}164}

				\addplot[
				mark = square*,
				thick,
				red,
				mark options = {solid},
				point meta = explicit,
				nodes near coords = {$n$=\pgfmathprintnumber[precision=0,1000 sep={,}]{\pgfplotspointmeta}},
				every node near coord/.append style = {font=\small, text=red, anchor=north west, yshift=-2pt, xshift=2pt}
				] table[
				x expr = {\thisrow{lower}+5},
				y = fraction_correct,
				meta = count
				] {\calibdsreason};
				\addlegendentry{DeepSeek (reasoning), $N$=1{,}153}
		\end{axis}
	\end{tikzpicture}
	\caption{Confidence calibration: fraction of correct verdicts per reported-confidence bin, pooled over the three full-dataset runs. Points sit at bin midpoints and carry the number of verdicts $n$ in the bin, and the dashed line shows perfect calibration. }
	\label{fig:confidence-calibration}
\end{figure}
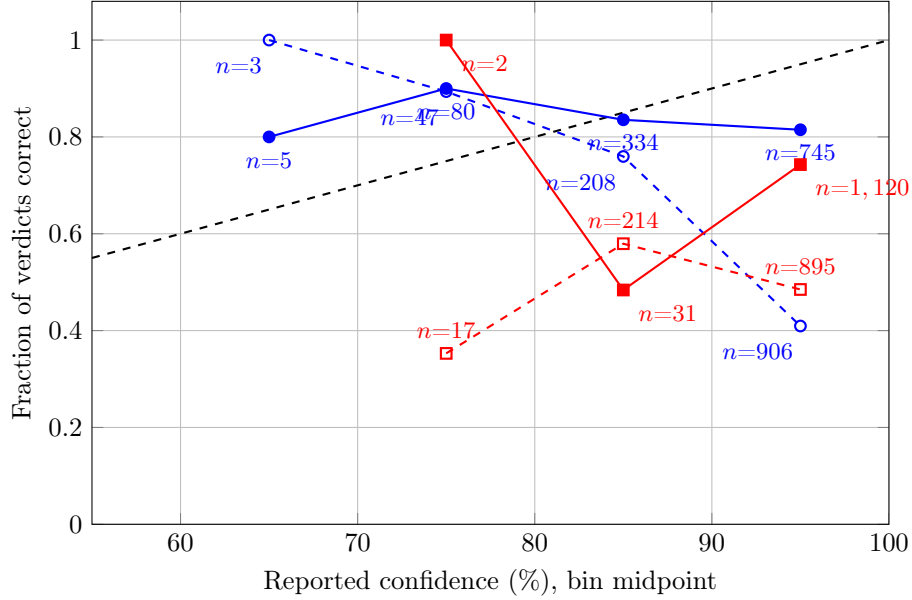

All models were instructed to report a confidence score between 1 and 100 for each verdict. Figure~\ref{fig:confidence-calibration} groups those scores into bins and plots the fraction of correct verdicts in each. Every series covers the same 388 goals in all three runs, so the four curves rest on the same 1{,}164 verdicts and their bin sizes show only how each configuration spreads its confidence. A few scores fall outside the bins. Thirty-eight DeepSeek chat and 11 DeepSeek reasoning verdicts reported exactly 0, a value the prompt's 1 to 100 scale does not permit, and we therefore treat these as missing rather than as self-reports. Both GPT series report a usable score on every verdict. The plotted totals are 1{,}164 for GPT chat, 1{,}126 for DeepSeek chat, 1{,}164 for GPT reasoning, and 1{,}153 for DeepSeek reasoning.

The reasoning models assigned uniformly high scores irrespective of correctness. Almost every verdict carried a confidence of 80 or above, and the large majority fell in the 90--100 band (GPT 745 of 1{,}164, DeepSeek 1{,}120 of 1{,}153). This overconfidence persists on incorrect verdicts. GPT rated its missed (\texttt{no attack}) verdict on NSPK at up to 95\% confidence despite the protocol's well-known authentication attack, and on the \texttt{Secret\_for\_B} family both models held 94--99\% confidence on attacks they failed to detect. Confidence carries almost no information about correctness. For GPT, accuracy in the top 90--100 bin (81.5\%) is slightly \emph{lower} than in the 70--79 bin (90.0\%), and DeepSeek's 90--100 bin is correct only 74.3\% of the time.

The chat models spread their scores more widely. For GPT chat the spread is inverted rather than merely uninformative. Its accuracy falls monotonically as reported confidence rises: 0.89 in the 70 to 79 bin, 0.76 in the 80 to 89 bin, and 0.41 in the 90 to 100 bin, which is where 906 of its 1{,}164 verdicts sit. The bin the model is most confident about is the one it gets wrong most often, and it is also the largest. DeepSeek chat shows no such ordering in either direction, moving 0.35, 0.58, 0.48 across the same three bins. Three bins are too thin to interpret, namely DeepSeek reasoning 70 to 79 with 2 verdicts, GPT reasoning 60 to 69 with 5, and GPT chat 60 to 69 with 3. Read across all four configurations, the proportion of correct verdicts never increases with reported confidence, the highest-confidence bins are not the most accurate, and many incorrect verdicts receive near-maximum confidence.

Self-reported confidence scores cannot be used as a filter to discard low-quality verdicts, as models exhibit overconfidence on precisely the outputs that are incorrect.

\subsection{Cost Analysis}
\label{sec:cost}
Table~\ref{tab:cost} splits the API spend into the chat and reasoning blocks. We exported the OpenAI figures from the provider dashboard and priced DeepSeek v4 Flash on the tokens consumed, so the two reasoning rows cover the same 390-request, three-run workload but each is priced on what that model consumed. The reasoning block dominates. GPT-5.4 answered 390 requests across 130 protocols over three runs, consuming 441,972 input tokens and 2,412,526 output tokens for \$37.29. Output tokens account for \$36.19 of that total, because a reasoning model bills its chain-of-thought as generated text, and that text far exceeds the prompt. DeepSeek v4 Flash, on the same workload, consumed 152,692 input and 650,669 output tokens per run, 803,361 in total. Over three runs that scales to 458,076 input and 1,952,007 output tokens, which at DeepSeek's per-token rates would have cost \$0.61, roughly sixty times less than GPT-5.4. Per protocol, the token total ranged from 1,645 (\texttt{ShareAgreeSimple}) to 66,669 (\texttt{NSL\_KeyServer}), with a mean of 6,179 and a median of 4,916.

The matched GPT arms let us price reasoning directly, because the only difference between them is whether the model reasons. Both arms sent an identical 441{,}972 prompt tokens over their three runs. The chat arm returned 352{,}141 completion tokens and reported no reasoning tokens at all. The reasoning arm returned 2{,}412{,}526, a factor of 6.85 more output for the same questions. At GPT-5.4's rates that is \$6.39 for the chat arm against \$37.29 for the reasoning arm, so enabling reasoning multiplied the bill on this workload by roughly six while lifting precision from 27.3\% to 64.8\%. Adding DeepSeek chat at \$0.20, the chat block comes to \$6.59 and the whole experiment to about \$44.49. The cost of a reasoning model lies almost entirely in the generated chain-of-thought output, and at scale that expense grows faster than the precision it provides.

\begin{table}[t]
	\centering
	\caption{API cost breakdown by experiment block. The GPT reasoning figure is the spend measured from the OpenAI dashboard. The GPT chat figure is priced on that arm's measured token consumption at the same rates. DeepSeek v4 Flash computes its own measured token consumption at DeepSeek's per-token rates during the experiment and saves the token consumption.}
	\label{tab:cost}
	\begin{tabular}{llr}
		\hline
		\emph{Experiment block} & \emph{Model(s)} & \emph{Cost (USD)} \\
		\hline
		Chat, full dataset, 3 runs      & GPT-5.4 (no reasoning) & 6.39 \\
		                                & DeepSeek v4 Flash (chat) & 0.20 \\
		Reasoning, full dataset, 3 runs & GPT-5.4            & 37.29 \\
		                                & DeepSeek v4 Flash  & 0.61 \\
		\hline
		\emph{Total} & & \emph{44.49} \\
		\hline
	\end{tabular}
\end{table}

\section{Discussion}
\label{sec:discussion}

\subsection{Reliability  of LLM-Generated Security Verdicts}
\label{sec:epistemic}

The results highlight a fundamental difference between formal verification tools and LLMs. Formal tools such as ProVerif and OFMC derive their conclusions from an explicit model, a set of inference rules, and a proof procedure that is sound with respect to that model. A verdict of \emph{no attack} within a given session bound, or even for an unbounded number of sessions, carries a logical guarantee relative to the underlying Dolev--Yao theory. LLM verdicts, by contrast, are the outcome of a probabilistic generation process conditioned on the prompt and on patterns absorbed during pre-training. They carry no such guarantee.

Our experimental findings reinforce this divide. Even the best-performing configuration, GPT reasoning on the full dataset, achieved only 66.5\% precision and 54.5\% recall against ProVerif, detecting barely more than half of the true attacks while still introducing false positives. The chat models, despite their higher recall, added many false alarms, with precision below 29\%. Accuracy does not improve the picture. On this imbalanced goal set, a predictor that answers \texttt{no attack} unconditionally scores 77.1\%, which only GPT reasoning exceeds, and then by 5.5 percentage points. The models are not competitive with formal verification on any framing of the metric, and the one framing on which they look respectable is the one that rewards saying nothing. The models' systematic failures on agreement goals, with low recall on both injective and non-injective agreement and no sign that they treat the two levels differently, suggest they lack a robust operational understanding of the authentication hierarchy. The confidence scores, uniformly high even for clearly incorrect decisions, show that the models have no reliable way to judge their own correctness.

The models’ sensitivity to multi-session attacks varies substantially across operational modes. When evaluated against the single-session OFMC benchmark, whose attack detection is limited, chat models flag many attacks that the bounded tool does not recognise, and a small proportion of these are confirmed by ProVerif or by OFMC at two sessions. This might suggest that LLMs can \emph{anticipate} cross-session vulnerabilities. However, the reasoning models, which are explicitly designed for more careful analysis, exhibit the opposite behaviour: they miss virtually all of the multi-session attacks that ProVerif identifies, effectively reverting to a single-session horizon. The divergence suggests that any multi-session sensitivity observed in chat mode comes from a lower threshold for issuing an attack verdict, not from principled reasoning about interleaved sessions. That threshold yields high recall at the expense of an unacceptable false-positive rate. The LLM is therefore not performing genuine symbolic exploration of concurrent traces. It is pattern-matching against structural features learned from training data, and those features correlate only weakly with the security of the protocol under the Dolev--Yao model.

In essence, LLM-generated security verdicts are fallible, data-driven heuristics, not sound logical proofs. They may capture surface-level regularities that sometimes align with genuine vulnerabilities, but they lack the soundness, completeness, and introspection that formal tools provide. Any practical deployment strategy must begin by acknowledging this unreliability.

\subsection{Limitations}
\label{sec:limitations}

While the experimental results offer a detailed empirical picture, a number of practical constraints limit the scope and certainty of the conclusions that can be drawn.

Both providers hold the model fixed across the two modes, so neither chat-versus-reasoning contrast is confounded by a change of model version: GPT runs \texttt{gpt-5.4} at reasoning effort \texttt{none} and \texttt{high}, and DeepSeek runs v4~Flash without and with its chain-of-thought. Each configuration was evaluated over three independent runs on the full 388-goal dataset. One caveat remains. Reasoning effort is a provider-controlled setting, and our evidence that it was honoured is the provider's own accounting, namely zero reported reasoning tokens on every chat-mode request. We cannot observe the model's internal computation directly, so a residual possibility remains that \texttt{effort: none} suppresses the emission of reasoning rather than the reasoning itself. The token evidence is consistent and the behavioural gap between the arms is large, but this rests on a documented API contract rather than on verified model internals.

\paragraph{Absence of manual expert review}
All LLM verdicts were compared against the formal-tool ground truth rather than against an expert reading of each protocol. Where an LLM and the ground truth disagree, the ground truth itself could in principle be incomplete. Three checks bound this risk.

First, the formal sources are highly consistent where they overlap (Section~\ref{sec:methodology}): ProVerif and OFMC agree on 94 to 98\% of jointly resolved goals, and essentially all disagreements are session-bound coverage effects rather than contradictions. Second, the definitive verdict is robust to its most uncertain component. Recomputing the reasoning-model metrics with the 18 ProVerif-uncovered goals dropped rather than filled by OFMC shifts precision and recall by at most about two percentage points: GPT moves from 64.8\% to 66.5\% precision and 53.2\% to 54.5\% recall, and DeepSeek from 44.4\% to 45.4\% and 55.1\% to 57.3\%.

Third, we examined the 18 ProVerif-uncovered goals directly, since these are the ones a reader would most reasonably distrust. They fall in ten protocols. Fourteen of the 18 are resolved by OFMC at two sessions, 4 of them attacks and 10 not, which leaves only \emph{four} goals resting on OFMC's single-session horizon. Those four are the sole place in the benchmark where a multi-session attack could be silently recorded as \emph{no attack}, and they are goals 2 and 3 of \texttt{Bullet\_Channel\_2\_Secure} and of \texttt{Bullet\_Channel\_3\_Secure}, all recorded \emph{no attack}. Three of the four were flagged as attacks by at least one configuration, so they are precisely the cases where model and ground truth disagree in the direction the ground truth cannot rule out. Four goals are 1.0\% of the 388, which bounds the exposure. Across the 18, every configuration agrees with the OFMC-supplied verdict on 13 and disagrees on 5. That rate is no worse than its performance on the ProVerif-resolved goals, so these goals are not anomalous.

The residual disagreements also cluster rather than scatter. The 23 goals that both reasoning models miss in all three runs are, without exception, agreement or secrecy goals: 15 injective agreement, 3 non-injective agreement, and 5 secrecy, spread over 15 protocols. Not one is a confidentiality or channel goal. A blind spot that respects goal-type boundaries this cleanly is more consistent with a systematic gap in how these models treat authentication than with noise in the benchmark. A full expert manual review of every disagreement would strengthen confidence further and remains future work. The targeted audit reported here is narrower, but it locates and bounds the specific exposure rather than leaving it unquantified.

\paragraph{Residual contamination risk} Type-aware identifier obfuscation (Section~\ref{subsec:obfuscation}) removes protocol names and identifier names, but it deliberately preserves the message structure, and that structure is itself recognisable. A model that has seen Needham--Schroeder in pre-training may still identify it from the shape and ordering of its messages once the names are stripped, so obfuscation lowers the contamination risk rather than eliminating it. We cannot bound the residual, and it plausibly differs across protocol families. Canonical textbook protocols are far more likely to be memorised than the corrected variants and industry-inspired designs in the same dataset. The decisive test is a set of synthetically generated protocols, or structurally modified ones unlikely to appear in any pre-training corpus, scored the same way. That experiment is future work. The same standard applies to our critique of prior benchmarks in Section~\ref{sec:related}. Obfuscation makes our estimate more conservative than one built on original names, not immune.

\paragraph{Generalisability} The study examined two LLM providers, OpenAI and DeepSeek, and one protocol specification language, \anb{}/\anbx{}. Two providers is a narrow base from which to generalise about ``current LLMs'', and the results may not transfer to other models, such as Anthropic's Claude Opus, or to other formalisms such as Tamarin’s multiset-rewriting rules. The 130-protocol dataset also covers a range of families and goal types without being exhaustive. Protocols involving complex algebraic properties, stateful sessions, or advanced equational theories are underrepresented. 
Furthermore, the wide confidence intervals obtained from the protocol-level cluster bootstrap (e.g., GPT precision 64.8\% (95\% CI:~[49.1, 78.1]) and recall 53.2\% (95\% CI:~[39.3, 66.3]) reflect the genuine variability in model performance across different protocol families. This uncertainty is not a flaw of the analysis but an honest quantification of how performance might change when applied to a different set of protocols.

\paragraph{Prompt and output constraints} We employed a single, zero-shot prompt template and forced a binary classification with no \emph{unknown} option. Holding the prompt fixed is what makes the chat-versus-reasoning contrast interpretable, but it comes at a price. Every number in this paper is conditional on that one prompt, and none should be read as a property of the models in general. Prompt sensitivity is well documented for LLMs, including in security tasks specifically \cite{white2023prompt,bae2024enhancing,nong2026}, so a prompt carrying more or less protocol-analysis guidance could plausibly move these figures. We therefore state our conclusions as conclusions about this prompt on these models. We did not establish that LLMs cannot analyse these protocols, only that they did not do so reliably under a carefully engineered zero-shot prompt of the kind a practitioner would most plausibly reach for first. Combined with the two-provider constraint above, this bounds the generalisability of the study in two independent directions. A prompt-variant ablation, holding the models and dataset fixed while varying the amount of analytical scaffolding, is the natural next experiment. The strict JSON schema, while necessary for automated evaluation, may also have constrained the models’ reasoning expressiveness.

\subsection{Implications for Practice}
\label{sec:implications}

Despite their current limitations, the results suggest a possible role for LLMs within a broader protocol-analysis workflow, provided their limited reliability is acknowledged. Verifying the entire 130-protocol suite with ProVerif takes less than an hour, and OFMC performs similarly for most protocols (though state explosion can prevent verification of large protocols with two parallel sessions). The results of these formal tools are also far more reliable than those of LLMs and incur no API cost. Nonetheless, the speed of LLM inference and their ability to handle arbitrary protocol sizes without state-space explosion make them potentially attractive for pre-screening very large or complex protocol collections where formal tools time out.

\paragraph{Pre-screening, not replacement} LLMs should not be deployed as autonomous verifiers. Chat-mode models offer the higher recall of the two modes, around 70 to 73\%, making them the better candidate for inexpensive first-pass triage. A large protocol library can be scanned quickly, with flagged cases prioritised for full verification. The limits of that role should be stated plainly. Roughly three in ten real attacks are still missed, so a chat-mode screen is a way of ordering verification effort, not a way of deciding what can safely skip verification. False positives also remain high, at close to three in four attack claims, so manual or tool-based follow-up is needed on everything the screen flags. Conversely, reasoning models achieve higher precision, suiting them for a secondary filtering step where fewer false alarms are desirable. However, their lower recall means many vulnerabilities would be missed, so reasoning models alone are not suitable for safety-critical filtering.

\paragraph{Inter-run disagreement as a signal} The observation that some protocols attract inconsistent verdicts across repeated runs provides an operational cue. If a majority-vote pipeline cannot reach a stable conclusion for a particular protocol, this very instability could indicate that the model is at the boundary of its reasoning capability and that the protocol should be escalated for formal analysis. However, further study is needed to confirm that it reliably correlates with protocol difficulty rather than stochastic noise.

\paragraph{Goal-type-aware strategies} The strong performance on confidentiality goals, with F1 up to 95.7\% for GPT reasoning, suggests that LLMs could be deployed with higher confidence for confidentiality screening. Secrecy goals sit well below that, at F1 53.6\% to 59.1\% in reasoning mode, so they warrant no comparable trust. For agreement goals, where all models detected well under half of the attacks, LLM analysis offers little added value and could be omitted from the pipeline entirely. A deployment strategy that tunes the reliance on LLM verdicts to the goal type would exploit the strengths of the models while mitigating their weaknesses. It is worth mentioning that formal tools are generally quicker to verify secrecy goals than authentication goals.

\paragraph{Cost-effectiveness} The cost analysis (Section~\ref{sec:cost}) shows that DeepSeek is roughly sixty times cheaper than OpenAI reasoning on the same workload, so it is the most plausible candidate for large-scale screening. The matched GPT arms sharpen the point. Turning reasoning off cuts that arm's cost by a factor of about six, from \$37.29 to \$6.39, because a reasoning model bills its chain-of-thought as generated output. The two chat configurations are statistically indistinguishable in accuracy (Section~\ref{sec:chat-results}), so the choice of provider for a first pass is a cost decision rather than a quality one, and DeepSeek chat is cheaper by an order of magnitude. A practical pipeline might use DeepSeek chat for that first pass, then DeepSeek reasoning or a formal tool on the flagged protocols, and keep OpenAI reasoning for the ambiguous cases that justify its cost.

\paragraph{Confidence scores must be ignored} Finally, the uniform failure of confidence calibration, with models assigning 94--99\% confidence to missed attacks, means that self-reported confidence should never be used to filter, weight, or rank LLM verdicts. Any workflow that incorporates LLM outputs must treat all verdicts as equally uncertain and subject to external validation.

\section{Related Work}
\label{sec:related}

\paragraph{LLMs in software security}
LLMs have attracted considerable attention in security, particularly for code-level vulnerability detection and repair. Sheng et al.~\cite{sheng2025llms} reviewed LLM-based vulnerability detection techniques and found that frontier models can achieve strong performance on common vulnerability classes, while still suffering from false positives and limited context awareness for complex repository-level dependencies. Yao et al.~\cite{Yao2024} surveyed LLMs across security and privacy sub-fields, documenting contributions to malicious code identification and automated patching, but cautioning that reliability remains insufficient for high-stakes deployment. Empirical studies have reinforced these concerns: Pearce et al.~\cite{Pearce2022} showed that GitHub Copilot frequently reproduces insecure code patterns, and Bae et al.~\cite{bae2024enhancing} demonstrated that vulnerability-detection performance varies materially with prompt style. Zhang et al.~\cite{Zhang2024} further observed that embedding structural or sequential information in prompts can improve LLM performance on vulnerability tasks, while Conceição et al.~\cite{Conceicao2025} concluded that models perform adequately on well-structured tasks with clear ground truth but remain prone to hallucination under adversarial conditions. Our work extends this line of inquiry from code-level security to symbolic protocol analysis, a domain where correctness can be formally specified and where the consequences of false negatives are comparably severe.

\paragraph{LLMs and formal methods for security}
Several recent efforts have investigated the combination of LLMs with formal verification tools for cryptographic protocols. The P2FGPT framework~\cite{li2025constructing} uses an LLM-based Generator–Checker–Modifier architecture to transform \anb{}-style protocol descriptions into ProVerif-compatible model declarations; the authors report reasonable accuracy but note that models can introduce new errors while fixing previous ones. Mao et al.~\cite{mao2025llm} proposed a staged pipeline that decomposes symbolic model generation into multiple steps to compensate for the unreliability of direct LLM output, achieving correct models in 10 out of 18 cases. CryptoFormalEval~\cite{curaba2024cryptoformaleval} integrates an LLM agent with the Tamarin prover, using Tamarin’s feedback to guide iterative analysis, but the architecture reflects an assumption that the LLM alone is insufficiently reliable. In all of these works, the LLM serves as an intelligent interface to, or a generator of input for, a sound formal tool; the correctness guarantee is ultimately deferred to the tool. Our study differs fundamentally: we treat the LLM itself as the verifier and compare its direct verdicts against formal ground truth, thereby providing a baseline measurement of LLM reasoning capability in isolation.
CrypFormBench~\cite{Li2026} introduced a large-scale benchmark for evaluating LLMs on formal cryptographic scheme analysis across seven verification languages and five tasks. Their work focuses on whether LLMs can generate tool-usable formal specifications from natural-language descriptions. However, their evaluation instances use original, non-obfuscated protocol names, introducing a risk of data contamination: LLMs may have encountered these exact specifications during pre-training, potentially inflating reported performance through memorisation rather than genuine reasoning. Our study addresses a fundamentally different question: whether LLMs can act as direct verifiers, producing security verdicts without relying on an external formal tool. Moreover, our use of type-aware identifier obfuscation (Section~\ref{subsec:obfuscation}) forces models to reason from the protocol's symbolic structure alone, providing a more conservative and honest assessment of their true analytical capabilities.

\paragraph{Prompt engineering for security tasks}
The quality of LLM outputs is highly dependent on prompt design, a finding that has been replicated across security-oriented applications. White et al.~\cite{white2023prompt} provide a general catalogue of prompt engineering patterns, emphasising the importance of explicit task framing and output format specification. In the vulnerability domain, Nong et al.~\cite{nong2026} report that chain-of-thought prompting improves analysis accuracy, while Wang et al.~\cite{wang2024chain} link the presence of explicit reasoning paths to increased user confidence in LLM-generated security assessments. Agarwal et al.~\cite{agarwal2024many} show that many-shot in-context learning can reduce the need for fine-tuning on complex reasoning tasks, albeit at the cost of increased token consumption. Nguyen et al.~\cite{nguyen2026thinking} advocate combining unconstrained natural-language reasoning with a later structured-generation phase to preserve reasoning richness while ensuring machine-parseable output. Our prompt design reflects these insights: we adopt a zero-shot structured prompt that embeds a comprehensive notation reference within a strict JSON output schema, avoiding in-context examples to minimise the risk of the model reproducing memorised solutions.

\paragraph{Uncertainty quantification in LLMs}
The unreliability of LLM outputs has motivated work on uncertainty quantification (UQ). Traditional UQ distinguishes aleatoric (noise/ambiguity) from epistemic (knowledge gaps) uncertainty. Liu et al.~\cite{liu2025uncertainty} propose a four-dimensional taxonomy for LLMs; Yu et al.~\cite{yu2024beyond} caution that binary decisions without an ``unknown'' option mask predictive uncertainty. Our study uses binary classification for straightforward metric computation but collects confidence scores for calibration analysis. The finding that these scores are uniformly high yet poorly calibrated adds to the evidence of LLM overconfidence in safety-critical domains.

\paragraph{Reasoning-oriented LLMs}
Recent reasoning-oriented LLMs incorporate explicit reasoning mechanisms: extended chain-of-thought, test-time computation, and reinforcement-learning optimisation. Surveys of mathematical reasoning~\cite{Wang2026} report substantial gains over conventional chat models on symbolic manipulation, theorem proving, and multi-step deduction, with examples such as OpenAI's o-series~\cite{openai2025o3} and DeepSeek-R1~\cite{deepseekReasoningDocs}. Whether these gains transfer to formal security reasoning remains unclear. Protocol verification differs from benchmark tasks: correctness depends on finding attack traces within a symbolic adversary model, not on producing plausible explanations. Our study directly compares chat and reasoning configurations on protocol-verification tasks with formal ground truth.

\paragraph{Position of the present study}
To the best of our knowledge, this is the first study to perform a direct comparison of LLM-generated security verdicts against formal verification ground truth, while simultaneously contrasting chat and reasoning model configurations across two independent providers. Unlike prior work that positions LLMs as assistants to formal tools, we evaluate the model as the primary analyst and assess its raw classification performance, its consistency across repeated runs, and the calibration of its self-reported confidence.

Methodologically, our study employs a rigorous type-aware identifier obfuscation pass that replaces all agent names, nonces, and keys with fresh, semantically neutral identifiers. This forces models to reason from the protocol's symbolic message structure rather than recognising well-known protocol names, a critical safeguard against training-data contamination. Through structured prompting and multi-metric evaluation (including goal-type disaggregation), we assess LLM capabilities and limitations in symbolic protocol verification.

\section{Conclusion}
\label{sec:conclusion}

This study asked whether large language models, in chat and reasoning configurations, can analyse security protocols specified in the \anb{}/\anbx{} notation as reliably as formal verification tools. The results, drawn from 130 protocols and 388 security goals scored against ProVerif and OFMC, lead to three principal conclusions.

First, the two modes fail in opposite ways. Chat models recall most true vulnerabilities, 72.7\% for GPT chat and 69.3\% for DeepSeek chat, but their precision stays below 29\%, so most of their attack claims are false alarms. Reasoning models reverse the trade-off. They reach higher precision, 66.5\% for GPT reasoning and 45.4\% for DeepSeek~v4~Flash against ProVerif, but recall falls to roughly 55\%, so they confirm only a slim majority of the real attacks. Each provider runs a single model in both modes, differing only in whether reasoning is enabled, so both precision gains are attributable to reasoning itself rather than to a change of model version. Neither configuration reaches the reliability expected of a formal verification tool. On this imbalanced goal set, neither chat configuration reaches even the 77.1\% accuracy of a predictor that answers \texttt{no attack} unconditionally.

Second, all models struggle with agreement goals. Reasoning models recover well under half of both injective and non-injective agreement attacks, with recall between 26 and 40\%, and the indiscriminate flagging of channel goals seen in chat mode gives way to erratic, low-volume handling in reasoning mode. The failures are systematic rather than incidental. Several protocols are missed by every model in every run, and the two reasoning models share 39 false negatives, concentrated in protocols that rely on a trusted third party or on multi-session interleaving. This points to a difficulty with concurrent session interactions. Verdicts do vary from run to run, but that variation does not account for the persistent core of missed attacks. Self-reported confidence scores are uniformly high, with almost every verdict carrying a confidence of 80 or above, and show no meaningful correlation with correctness, which makes them useless as a reliability filter.

Third, these limits still leave room for a narrow practical role. Because chat models recall most attacks, they can serve as a first-pass triage filter that flags which protocols warrant formal verification, and because reasoning models are more precise, they can help confirm candidate attacks with fewer false alarms. Cost reinforces that division of labour. DeepSeek is roughly sixty times cheaper than OpenAI reasoning on the same workload, which makes it the more plausible candidate for the high-volume first pass. Such a workflow has to stay goal-type-aware. It should lean on the strong confidentiality results and avoid trusting LLM verdicts on injective agreement, where recall stays below 41\%. It should also treat disagreement between runs as a candidate trigger for escalation. Verdicts are identical across runs on 89.7\% of goals for GPT reasoning, falling through 74.0\% for DeepSeek reasoning and 70.1\% for GPT chat to 61.6\% for DeepSeek chat. Aggregating three runs by majority vote recovers part of that gap in reasoning mode, raising DeepSeek precision from 44.4\% to 47.1\%, but recovers nothing in chat mode, where the errors are systematic rather than random. No LLM verdict should be accepted without independent formal confirmation.

Future work should explore several directions. \emph{Multi-model comparison} is needed: applying the same protocol corpus and evaluation methodology to models from other families, such as Claude and Gemini, would reveal whether the limitations documented here are specific to the tested providers or represent a general capability boundary for the current generation of LLMs. \emph{Extending the benchmark} to include protocols with complex algebraic properties and stateful sessions would test the generality of our findings and provide a more demanding evaluation framework for future LLM-based protocol analysers. \emph{Validating the attack traces} that models emit, by mechanically replaying each reported \texttt{two\_session\_trace} in a tool such as OFMC, would separate genuine attack discovery from spurious narrations. \emph{A prompt-variant ablation} would test how far these results depend on the single zero-shot prompt used throughout. Varying the amount of protocol-analysis guidance, while holding the models and dataset fixed, would show whether the failure modes documented here are properties of the models or of the way we asked. \emph{Synthetically generated protocols}, or structurally modified ones unlikely to appear in any pre-training corpus, would close the residual contamination gap that identifier obfuscation alone cannot (Section~\ref{sec:limitations}).

\bibliographystyle{elsarticle-num}
\bibliography{literature}

\appendix

\section*{Abbreviations}
\vspace{-1.5\baselineskip}
\setlength{\nomitemsep}{-\parsep}
\renewcommand{\nomlabel}[1]{\textbf{#1}} 
\nomenclature{AI}{Artificial Intelligence}
\nomenclature{AnB}{Alice and Bob notation}
\nomenclature{AnBx}{Extended Alice and Bob notation}
\nomenclature{API}{Application Programming Interface}
\nomenclature{AVISPA}{Automated Validation of Internet Security Protocols and Applications}
\nomenclature{CoT}{Chain-of-Thought}
\nomenclature{CSV}{Comma-Separated Values}
\nomenclature{DH}{Diffie--Hellman}
\nomenclature{DY}{Dolev--Yao}
\nomenclature{EMV}{Europay, Mastercard, and Visa (smart payment standards)}
\nomenclature{FN}{False Negative}
\nomenclature{FP}{False Positive}
\nomenclature{F1}{Harmonic mean of precision and recall}
\nomenclature{GPT}{Generative Pre-trained Transformer}
\nomenclature{HMAC}{Hash-Based Message Authentication Code}
\nomenclature{iKP}{i Key Payment protocol family}
\nomenclature{JSON}{JavaScript Object Notation}
\nomenclature{LLM}{Large Language Model}
\nomenclature{MD5}{Message Digest Algorithm 5}
\nomenclature{NSPK}{Needham--Schroeder Public-Key protocol}
\nomenclature{OFMC}{On-the-Fly Model Checker}
\nomenclature{SHA}{Secure Hash Algorithm}
\nomenclature{SET}{Secure Electronic Transaction}
\nomenclature{TLS}{Transport Layer Security}
\nomenclature{TN}{True Negative}
\nomenclature{TP}{True Positive}
\nomenclature{UQ}{Uncertainty Quantification}

\printnomenclature[20mm]

\section{Prompt Used for Protocol Analysis}\label{app:prompt}

\begin{lstlisting}[
style=mystyle,
numbers=none,
basicstyle=\ttfamily\footnotesize,
breaklines=true,
breakatwhitespace=true,
columns=fullflexible,
frame=single,
caption={Full prompt used for LLM protocol analysis},
label={lst:full-prompt-template}
]
PROMPT_TEMPLATE = """
You are a security protocol verifier using symbolic reasoning.

You are given:
1) an anonymized Alice-and-Bob / AnB / AnBx protocol body, and
2) an explicit list of goals to analyze.

--- NOTATION REFERENCE ---
Use the following Alice-and-Bob notation when interpreting the protocol:

Keys:
  pk(X)        : public key of X used for encryption
  sk(X)        : public key of X used for signature verification
  inv(pk(X))   : private key of X used for decryption
  inv(sk(X))   : private key of X used for signature
  shk(X,Y)     : pre-shared symmetric key between X and Y

Encryption:
  {|Msg|}K        : symmetric encryption of Msg under key K (e.g. AES, DES)
  {Msg}pk(A)      : asymmetric encryption of Msg under public key of A (e.g. RSA)
  {Msg}inv(sk(A)) : digital signature of Msg using private key of A (e.g. RSA, DSA)

Where:
  Msg is a message, K is a symmetric key, A is an agent

Hashing:
  hash(Msg)    : cryptographic hash of Msg (e.g. MD5, SHA-1, SHA-2, SHA-3)
  hmac(K, Msg) : keyed HMAC of Msg under symmetric key K (e.g. HMAC-SHA1, HMAC-SHA2)

--- CHANNELS ---

A -> B: Msg   : Message exchange on an insecure channel
A ->* B: Msg  : Message exchange on a confidential channel
A *-> B: Msg  : Message exchange on an authenticated channel
A *->* B: Msg : Message exchange on a secure channel

--- GOAL DEFINITIONS ---
Authentication (hierarchy, weakest to strongest):
  1. Non-injective Agreement: Adds the requirement that A and B agree on specific data items (such as nonces and keys) and the roles they played.
  2. Injective Agreement: The strongest form -- adds a one-to-one relationship between the runs of A and B, preventing replay attacks where B believes multiple runs occurred corresponding to a single run by A. Equivalent to what OFMC/ProVerif verify as injective agreement.

Secrecy:
  Confidentiality: An attacker cannot derive a protected term from intercepted messages. (e.g. A confidentially sends Msg to B, or A ->* B: Msg).
  Secrecy: An attacker cannot distinguish between protocol executions that differ only by their secret inputs (e.g. Msg secret between A,B)

--- RULES ---
- Analyze ONLY the listed goals. Never invent extra goals.
- Use Dolev-Yao attacker assumptions (network control, no cryptographic breaks without keys).
- For EVERY goal, output:
  - status: "attack found" or "no attack"
  - confidence: integer from 1 to 100 reflecting how certain you are in that verdict
  - justification: 2 to 6 sentences
  - two_session_trace: always present, exactly two sessions (Session 1 and Session 2)
- Do not mention filenames, folders, or real protocol names. Refer only to protocol_id.

Return JSON only. No markdown.
Schema:
{{{{
  "model": "<string>",
  "protocol_id": "<string>",
  "analysis": [
    {{{{
      "goal_id": <int>,
      "goal": "<goal text>",
      "status": "attack found|no attack",
      "confidence": <int 1-100>,
      "justification": "<100 words>",
      "two_session_trace": "Session 1: ...\\nSession 2: ..."
    }}}}
  ]
}}}}

PROTOCOL_ID:
{protocol_id}

PROTOCOL_BODY:
{protocol_body}

GOALS:
{goals_block}
""".strip()
\end{lstlisting}

Listing~\ref{lst:full-prompt-template} presents the full prompt used for protocol analysis. The prompt combined notation guidance, attacker assumption, goal definitions, and a constrained JSON output schema in order to reduce ambiguity and ensure that model outputs could be parsed automatically for evaluation.

\end{document}